\documentclass[aps,pre,preprint,groupedaddress,showkeys]{revtex4-1}

\usepackage{color}
\usepackage{amsfonts}
\usepackage{amsmath}
\usepackage{graphicx}
\newtheorem{theorem}{Theorem}
\newtheorem{lemma}{Lemma}

\newcommand{\Ei}{\mbox{\rm Ei}}
\newcommand{\newtext}[1]{\textcolor{blue}{#1}}
\begin{document}


\title{
Exact asymptotic solution of an aggregation model 
with a bell-shaped distribution
}


\author{F. Leyvraz}
\email[]{leyvraz@icf.unam.mx}
\altaffiliation{Centro Internacional de Ciencias, Cuernavaca, Morelos, M\'exico}
\affiliation{Instituto de Ciencias F\'\i sicas---Universidad Nacional Aut\'onoma de 
 M\'exico, Cuernavaca, Morelos, M\'exico}


\date{\today}

\begin{abstract}
We present in a detailed manner the scaling theory of  irreversible aggregation 
characterized by the set of reaction rates $K(k,l)=1/k+1/l$. In this case, it is possible 
to  determine the behaviour of large-size aggregates in the limit of large times
in a way that allows a highly detailed analysis of the behaviour of the system. This is the so-called 
scaling limit, in which the cluster size distribution collapses to a function of the ratio of the cluster
size to a time-dependent {\em typical size}. 
The results confirm the far more general results of earlier work concerning
a general scaling theory for so-called reaction rates of Type III,  which are characterised 
by the property that aggregates of very different sizes react faster than comparable aggregates of
similar sizes. For these, the cluster size distribution decays rapidly to zero both for sizes 
much larger and much smaller than the typical size,
and is thus often described as being ``bell-shaped''. For clusters much larger than the typical size, however,
an unexpected subleading correction is discovered. 
 Finally, several results going beyond the scope of
the scaling limit are obtained: in particular the behaviour of concentrations for fixed cluster size in 
the large-time limit and the large-size behaviour for clusters at a fixed time. The latter again shows 
subleading deviations from the expected behaviour.
\end{abstract}


\maketitle

\section{Introduction}

Irreversible aggregation is the process whereby aggregates grow by the scheme
\begin{equation}
A_k+A_l\mathop{\longrightarrow}_{K(k,l)}A_{k+l},
\end{equation}
with no backward reaction. Here $A_k$ denotes an aggregate consisting of $k$ elemental aggregates (monomers)
and $K(k,l)$ denotes the dependence of the rate at which the aggregation occurs, as
a function of the masses of the 2 aggregates. 

Such processes occur in a broad variety of physical systems. Thus in aerosols and colloids 
these are quite common, as well as in many other contexts. The concentration
$c_j(t)$ of aggregates of mass $j$ varies according to the following
kinetic equation
\begin{equation}
\dot{c}_j(t)=\frac12\sum_{k,l=1}^\infty K(k,l)c_k(t)c_l(t)\left[
\delta_{j,k+l}-\delta_{j,k}-\delta_{j,l}
\right].
\label{eq:basic}
\end{equation}
The size dependence of the $K(k,l)$ is determined by the detailed physics of the
process under study. For aerosols which diffuse and coalesce irreversibly
in three dimensions, for instance,
a standard approximation for the rate is given by
\begin{equation}
K(k,l)=\left[
R(k)+R(l)
\right]\left[
D(k)+D(l)
\right]
\label{eq:diff}
\end{equation}
where $R(k)$ describes the typical radius of an aggregate of mass $k$ and $D(k)$
its diffusion constant  (for greater background on the physics of such systems, see for 
example \cite{smoke,drake,prupp}). If the aggregates are spherical objects, we have $R(k)$ proportional
to $k^{1/3}$ and similarly $D(k)$ goes as $k^{-1/3}$.

At this point it may be useful to point out that irreversible aggregation as treated here is only 
a small part of a large field of research, so rich that attempting to sketch it would be idle. 
Suffice it to say that, even for pure aggregation, it is possible to consider such variants as the
introduction of source terms, multiple species, spatial structure and the interaction with diffusion, as
well as many other issues. Some of these variants are described in \cite{ley03}. On the other hand, we may also
consider fragmentation, or fragmentation in coexistence with aggregation. The review \cite{sinica} together with the references therein
may be useful to the reader interested in this subject. Finally, for a broad view of related topics, \cite{imposed} is of 
considerable interest, whereas for applications in concrete systems, see \cite{smoke,drake,prupp}.

A general scaling theory for the solutions of (\ref{eq:basic}) was developed by \cite{vDongen2,vDongen3,vDongen1}
following initial earlier work in \cite{fried66},
and reviewed in \cite{ley03}. In it, it is assumed that there exist a function $S(t)$ diverging as $t\to\infty$,
which corresponds to the typical size of the aggregate size distribution at time $t$, and a function $\Phi(x)$
such that
\begin{equation}
c_j(t)\simeq \left[S(t)\right]^{-2}\Phi
\left(
\frac{j}{S(t)}
\right).
\end{equation}
 Here the symbol $\simeq$
simply signifies that the two sides of the equation grow at approximately the same rate 
in the appropriate limit. 
I specifically do not wish to commit myself 
to any specific mathematical statement when using this notation. 

The scaling theory makes two kinds of predictions. One is extremely general, and concerns the 
rate of growth of $S(t)$. Let the reaction kernel $K(k,l)$ be homogeneous of degree $\lambda$, that is
\begin{equation}
K(ak, al)=a^\lambda K(k,l)
\end{equation}
at least asymptotically as $k,l\to\infty$. Then the typical size $S(t)$ grows as
\begin{equation}
S(t)\simeq t^{1/(1-\lambda)}\qquad(t\to\infty).
\end{equation}
A more detailed analysis concerns the small-$x$ behaviour of $\Phi(x)$, which 
leads to the distribution of cluster sizes for sizes much less than the typical size. 

To this end, we need to define another exponent characterising the behaviour
of $K(k,l)$ for very different values of $k$ and $l$: first let us
define
\begin{equation}
K(k,l)=k^\lambda {\cal K}\left(
\frac{l}{k}
\right).
\end{equation}
The exponents $\mu$ and $\nu$ are then given by
\begin{subequations}
\begin{eqnarray}
&&{\cal K}(x)\simeq x^{\nu}\qquad(x\to\infty),\\
&&{\cal K}(x)\simeq x^{\mu}\qquad(x\to0).
\end{eqnarray}
\end{subequations}
It follows immediately from the definition that
\begin{equation}
\mu+\nu=\lambda
\end{equation}
For example, if we take the case of the rates $K(k,l)$ given by
(\ref{eq:diff}):
\begin{equation}
K(k,l)=(k^{1/3}+l^{1/3})(k^{-1/3}+l^{-1/3}),
\label{eq:diff1}
\end{equation}
we find the values $\lambda=0$, $\mu=-1/3$ and $\nu=1/3$. 

These exponents define to a large extent the nature of the scaling behaviour of the
system. First, all our considerations pertain to the case $\lambda<1$. When
$\lambda>1$, a divergence of the typical size followed by loss of mass to an infinite 
aggregate occurs at finite time. Rates for $\lambda<1$ are 
divided in 3 broad types, according to the sign of $\mu$  \cite{vDongen2}. 
Type I corresponds to $\mu>0$, Type II to $\mu=0$, and Type III 
to $\mu<0$ .The kernel (\ref{eq:diff1}) is thus of Type III. 

The scaling function behaves for $x\ll1$ differently for the 3 types. For type I we
have 
\begin{equation}
\Phi(x)\simeq x^{-\tau},\qquad\tau=1+\lambda
\end{equation}
for $x\ll1$. This means that there is a broad range of sizes over which the aggregates
are power-law distributed. On the other hand, Type II kernels behave in a quite non-universal manner.
In most known cases, $\Phi(x)$ also goes as a power $x^{-\tau}$, but for a value 
of $\tau$ which depends on the detailed behaviour of $K(k,l)$. Finally, Type III
kernels have a scaling function $\Phi(x)$ which goes faster than any power
towards zero as $x\to0$. Specifically
\begin{equation}
\Phi(x)\simeq const.\cdot x^{-\lambda}\exp\left(
-const.\cdot x^{-|\mu|}
\right).
\end{equation}
The aggregate sizes do not show a great variation and their
distribution is sometimes described as being bell-shaped.

Another issue is the behaviour of $\Phi(x)$ for $x\gg1$  which determines the behaviour
of the concentrations of clusters of size much larger than the typical size. In that case, 
$\Phi(x)$ always decays exponentially, but also has a correction exponent $\theta$
defined by
\begin{equation}
\Phi(x)\simeq const.\cdot x^{-\theta}\exp\left(
-const.\cdot x
\right).
\end{equation}
For all kernels we have
\begin{equation}
\theta=\lambda,
\end{equation}
except in the case $\nu=1$, for which a similar lack of universality exists as 
for the $x\ll1$ behaviour when $\mu=0$ \cite{vDongen3}.

There are several rate kernels for which (\ref{eq:basic}) can be solved exactly.
Interestingly, if we limit ourselves to the non-gelling case
$\lambda\leq1$, they are all of type II; the most prominent examples are
\begin{subequations}
\begin{eqnarray}
K_1(k,l)&=&\alpha+\beta(k+l),
\label{eq:exacta}
\\
K_2(k,l)&=&2-q^k-q^l\qquad(q<1),
\label{eq:exactb}
\\
K_3(k,l)&=&\alpha\delta_{k,1}\delta_{l,1}+\beta\left(
\delta_{k,1}+\delta_{l,1}\right)+\gamma,
\label{eq:exactc}
\\
K_4(k,l)&=&\alpha+\beta\left[
(-1)^k+(-1)^l
\right]+\gamma(-1)^{k+l}.
\label{eq:exactd}
\end{eqnarray}
\label{eq:exact}
\end{subequations}
where $\alpha$, $\beta$, and $\gamma$ are arbitrary constants, positive or zero \cite{spouge, cal99,mob03,cal00}. 
It would thus be of interest
to obtain any kind of rigorous information on kernels of type I or III. In particular, 
it is of interest to obtain exact, or at least accurate, asymptotic results, for a kernel 
of type III, since these are the kind that arises in the most natural way in aerosol physics.
 This is what we shall do in this paper.

In the following we consider the case, also treated earlier in \cite{mim09}, given by
\begin{equation}
K(k,l)=\frac1k+\frac1l.
\label{eq:ker1}
\end{equation}
 It is also straightforward to show that
the faintly more general version
\begin{equation}
K(k,l)=\frac1{\alpha k+\beta}+\frac1{\alpha l+\beta },
\label{eq:ker2}
\end{equation}
 where $\alpha\geq0$ and $\beta>-\alpha$ are arbitrary,  behaves in the scaling limit
in exactly the same way as the kernel given by (\ref{eq:ker1}), which shows that the behaviour we describe is at least
to some extent universal. We shall not have anything else to say about this generalised kernel, however. 

 The kernel (\ref{eq:ker1}) has $\lambda=\mu=-1$, so that we have
\begin{equation}
S(t)\simeq \sqrt t.
\end{equation}
and presumably
\begin{equation}
\Phi(x)\simeq const.\cdot x^{-1}\cdot\exp\left(
-const.\cdot x^{-1}
\right)\qquad(x\ll1).
\end{equation}
On the other hand, for $x\gg1$, we obtain
\begin{equation}
\Phi(x)\simeq const.\cdot x\exp\left(
-const.\cdot x
\right).
\end{equation}
In the following we shall show rigorously the above relations, and additionally 
evaluate explicitly the different constants involved. We further provide
a technique to evaluate the scaling function to high accuracy. We shall also obtain
subleading corrections to the behaviours stated above. 

Beyond the scaling limit, some other issues can also be treated. In particular we discuss 
the large-time behaviour of clusters of fixed size, which cannot be found via scaling. Similarly, the 
limit of large sizes for clusters at a fixed given time can be treated to a high degree of accuracy. 
In earlier treatments \cite{vDongen4,theta}, this behaviour was obtained through an approximation
valid only in the limit of small times. Here we solve exactly this small-time limit for the reaction rates 
(\ref{eq:ker1}). We find it to involve an exponential decay
modified by a power-law prefactor. Both the large-size limit of the scaling function
and the finite-time large size behaviour differ from the small-time behaviour
by a non-trivial correction to the power-law prefactor.

In Section \ref{sec:notation} we introduce the basic techniques as well as the notation. In Section
\ref{sec:summ} we summarise the results to be developed. In Section \ref{sec:deriv} we derive the results
for the scaling behaviour summarised in Section \ref{sec:summ}. In Section \ref{sec:largesizes}
and in Section \ref{sec:largetimes} we derive results beyond the scaling approximation: in the former,
we describe the behaviour of large clusters at fixed times, and in the latter we analyse 
the behaviour of cluster of given size at large times. Finally we present conclusions in Section
\ref{sec:conclusions}. Several Appendices clear up various technical points.
In particular, Appendix \ref{app:a} presents the detailed properties of a solution of a given nonlinear ODE,
which is closely related to the scaling function of the problem. 

\section{Problem and fundamental equations}
\label{sec:notation}
The  Smoluchowski's equations (\ref{eq:basic}) read, for the case we are interested in
\begin{equation}
\dot c_j=\sum_{k=1}^{j-1}\frac{c_kc_{j-k}}{k}-\frac{c_j}{j}\sum_{k=1}^\infty c_k-
 c_j\sum_{k=1}^\infty \frac{c_k}{k}  
\label{eq:1.2}
\end{equation}
Kernels of the form
\begin{equation}
K(k,l)=f(k)+f(l)
\label{eq:1.3}
\end{equation}
can be treated using a standard transformation, introduced by Lushnikov \cite{lush73}.
We introduce the new dependent and independent variables
\begin{subequations}
\label{eq:1.4}
\begin{eqnarray}
N(t)&=&\sum_{k=1}^\infty c_k(t)\label{eq:1.4a}\\
\phi_j&=&\frac{c_j}{N(t)}\label{eq:1.4b}\\
d{s}&=&N(t)dt.\label{eq:1.4c}
\end{eqnarray}
\end{subequations}
In these  new variables, the equations (\ref{eq:1.2}) read
\begin{equation}
\frac{d\phi_j}{d{s}}=\sum_{k=1}^{j-1}\frac{\phi_k\phi_{j-k}}{k}-\frac{\phi_j}{j}
\label{eq:1.5}
\end{equation}
This can be carried through for any kernel of the form (\ref{eq:1.3}) and eliminates the
non-recursive removal terms, which is always a considerable step towards
the solution. Indeed, it is clear that an exact expression for $\phi_j({s})$ can always be
obtained recursively, by solving a linear equation with a (recursively known)
inhomogeneity. The rapidly growing complexity of the resulting expressions
makes this approach impractical. As an example, the first 4  explicit 
expressions for $\phi_j(s)$ are given by:
\begin{subequations}
\begin{eqnarray}
\phi_1({s})&=&e^{-{s}}
\\
\phi_2({s})&=&\frac{2e^{-2{s}} }{3} 
\left(e^{3 {s}/2}-1\right)
\\
\phi_3({s})&=&\frac{ e^{-3 {s}}}{56}
 \left(-48 e^{3 {s} /2}+27
 e^{8 {s} /3}+21\right)\\
\phi_4({s})&=&\frac{2
e^{-4 {s}} }{36\,855} 
\big(13\,000 
e^{3{s} /2}-10\,935 e^{8 {s} /3}-5\,460
e^{3 {s}}+\nonumber\\
&&\qquad 6\,944 e^{15 {s}/4}-3\,549\big)
\end{eqnarray}
\label{eq:phirec}
\end{subequations}
 However, no clear pattern appears to be recognisable.
In the present case additional progress can be made: define the generating function
\begin{equation}
G(z,{s}):=\sum_{j=1}^\infty\frac{\phi_j({s})}{j}e^{-jz}. 
\label{eq:1.6}
\end{equation}
(\ref{eq:1.5}) then reads
\begin{equation}
G_{z{s}}(z, {s})=
G_z(z, {s})
G(z,{s})+G(z, {s}),
\label{eq:1.7}
\end{equation}
 where the subscripts indicate a partial derivative with respect to the variable.
Frequently, being able to cast a  problem as a PDE in the way we have
done here leads straightforwardly to the full solution. This is not the case
here: the PDE (\ref{eq:1.7}) is surprisingly complex, and I have made no 
headway at all.

Let us then look at the scaling limit. Using the standard approach discussed
for example in \cite{ley03}, the scaling theory predicts the existence of a function
$S(t)$ which grows as $\sqrt t$ and of a function $\Phi(x)$ such that 
\begin{equation}
\lim_{t\to\infty}\sum_{j=1}^\infty
jc_j(t)f\left(
\frac j{S(t)}
\right)=\int_0^\infty x\,\Phi(x)f(x)dx
\label{eq:1.8}
\end{equation}
holds for all appropriate functions $f(x)$. Putting $f(x)=1/x$---which is 
possibly a problem from a rigorous viewpoint, but we are now only 
making plausibility considerations---one immediately  obtains 
that 
\begin{equation}
N(t)=const.\cdot S(t)^{-1}=O(t^{-1/2}).
\end{equation}
From this follows  that we expect the
following scaling form for $G(z, {s})$:
\begin{equation}
G(z, {s})=\frac1{s}\Psi(z{s}).
\label{eq:1.9}
\end{equation}
Putting this into (\ref{eq:1.7}) leads to
\begin{equation}
\rho\Psi^{\prime\prime}=\Psi\Psi^\prime+\Psi.
\label{eq:1.10}
\end{equation}
 Here we may note that, if we had carried out an exactly identical computation using
the general kernel (\ref{eq:ker2}), we would also have obtained (\ref{eq:1.10}), thus justifying the remark that the
two kernels are rigorously identical in the scaling limit. Again I  have not been able to find a solution of this apparently simple equation.
In Appendix \ref{app:a} I show a large number of detailed properties of (\ref{eq:1.10}). In particular,
I show that there is a unique solution such that $\Psi(\rho)$ is positive, smooth
at the origin and monotonically decreasing to zero. Solutions smooth at the origin
are characterised by the initial conditions
\begin{subequations}
\begin{eqnarray}
\Psi(0)&=&1,
\label{eq:inita}
\\
\Psi^\prime(0)&=&-1,
\label{eq:initb}
\\
\Psi^{\prime\prime}(0)&=&\kappa,
\label{eq:initc}
\end{eqnarray}
\label{eq:init}
\end{subequations}
where $\kappa$ is arbitrary. The unique monotonically decreasing positive solution
is characterised by a uniquely defined value $\kappa_0$ 
numerically found to be between $1.45582$ and $1.45583$. All other solutions of the 
type described by (\ref{eq:init}) diverge (for $\kappa>\kappa_0$) or become negative
(for $\kappa<\kappa_0$). Since these
behaviours cannot arise in the function we are looking for, it is clear 
that the unique solution defined by $\kappa=\kappa_0$ is the only acceptable one. 

The fact that the solution of (\ref{eq:1.2}) actually tends towards the scaling solution defined 
by (\ref{eq:1.10}) follows from the work of Norris \cite{norris}, whereas the fact that the 
integro-differential equation which the scaling function can be shown to satisfy
\cite{vDongen1} actually has a solution under fairly general circumstances, was shown
in \cite{four05}. The peculiar feature of our approach, however, resides in the very explicit
nature of the equations to be solved, and the rather immediate, though somewhat tedious, 
character of the proof of existence, which is entirely constructive. It is performed 
in detail in Appendix \ref{app:a}. We also show in Appendix \ref{app:f} in detail that (\ref{eq:1.10})
follows from the well-known integro-differential equation introduced in \cite{vDongen1,fried66}.

\section{Summary of results}
\label{sec:summ}
 
Since the results presented are rather numerous, I wish to summarise them in the following for the reader's
convenience. But first a matter of notation. Above I have used the symbol $\simeq$ to denote a very rough concept of
similar growth, with no particular commitment to any specific meaning.  On the other hand, when I wish to make a
more specific statement, for instance that $f(t)$ and $g(t)$ approach a common limit, I shall express this as
\begin{equation}
\lim_{t\to\infty}\frac{f(t)}{g(t)}=1,
\end{equation}
or else, using the $o$-notation of Landau
\begin{equation}
f(t)=g(t)\left[
1+o(1)
\right]
\end{equation}
Here $o(1)$ represents a quantity which tends to zero as $t\to\infty$. Since we shall be interested in the
precise rate of growth of various quantities, these notations will be useful.

Essential to the entire further development is the following connection between  $\Psi(\rho)$, the 
solution of (\ref{eq:1.10}) described in Appendix \ref{app:a}, and the scaling function $\Phi(x)$
defined in (\ref{eq:1.8})
\begin{equation}
\Psi(\rho )=\kappa_0
\int_0^\infty dx\frac{\Phi(x)}{x}e^{-\rho x}.
\label{eq:psiscalsumm}
\end{equation}

\subsection{Large-time behaviour of moments}
 
We obtain the following rigorous results on the large-time behaviour of the moments of
$c_j(t)$ for $n\geq0$: let
\begin{equation}
\mu_n(t)=\sum_{j=1}^\infty j^n c_j(t).
\end{equation}
The asymptotic behaviour of $\mu_n(t)$ is then given by, in terms of the variable $s$ instead 
of $t$:
\begin{subequations}
\begin{eqnarray}
\mu_n({s})&=&m_n^{(\infty)}{s}^{n-1}\left[
1+o(1)
\right]\qquad(s\to\infty),
\label{eq:moma}\\
m_n^{(\infty)}&=&\kappa_0^{-1}(-1)^{n+1}\left.
\partial_\rho^{n+1}\Psi(\rho)
\right|_{\rho=0}.
\label{eq:momb}
\end{eqnarray}
\label{eq:mom}
\end{subequations}
Explicit expressions for $m_n^{(\infty)}$ are obtained from (\ref{eq:momb}) by recursive evaluation
of the derivatives of $\Psi(\rho)$ at the origin by the techniques described in Appendix \ref{app:a}, see in particular
(\ref{eq:a2.3}). 
We obtain for instance
\begin{subequations}
\begin{eqnarray}
m_0^{(\infty)}&=&\frac1{\kappa_0}
\label{eq:varmom0}\\
m_1^{(\infty)}&=&1
\label{eq:varmom1}\\
m_2^{(\infty)}&=&2
\label{eq:varmom2}\\
m_3^{(\infty)}&=&3\left(
1+\frac{\kappa_0}2
\right)
\label{eq:varmom3}
\end{eqnarray}
\label{eq:varmom}
\end{subequations}
Finally, expressions for $n=-2$ and $-1$ are also found
\begin{subequations}
\begin{eqnarray}
m_{-1}^{(\infty)}&=&\frac1{\kappa_0}
\label{eq:negmoma}
\\
m_{-2}^{(\infty)}&=&\frac3{2\kappa_0}
\label{eq:negmomb}
\end{eqnarray}
\label{eq:negmom}
\end{subequations}
The connection between $s$ and $t$ is also found asymptotically:
\begin{equation}
\lim_{{s}\to\infty}\frac{2t({s})}{\kappa_0{s}^2}=1.
\label{eq:taut}
\end{equation}
In the large-$n$ limit, the $m_n^{(\infty)}$ behave as
\begin{equation}
m_n^{(\infty)}=\frac{2}{\kappa_0}(n+1)!\Lambda^{-(n+1)}
\left[1-\frac{2\Lambda}{3n(n+1)}
\right]
\left[1+o(1)
\right]
\qquad(n\to\infty). 
\label{eq:momasym}
\end{equation}
Here $-\Lambda$ is the nearest singularity of $\Psi(\rho)$ to the origin, which, as shown 
in Appendix \ref{app:a}, lies on the negative real axis. It is numerically evaluated in Appendix
\ref{app:b}.

\subsection{Behaviour of the scaling function $\Phi(x)$ for small and large clusters}

The scaling function defined by (\ref{eq:1.8}) for $x\to0$ behaves as
\begin{equation}
\Phi(x)=\frac{\Gamma}{\kappa_0}\frac{\exp(-1/x)}{x}\left[
1+o(1)
\right]+const.\cdot \frac{\exp(-4/x)}{x^{7/2}}\left[
1+o(1)
\right].
\label{eq:phismall}
\end{equation}
Here $-\Lambda$ is, as above, the nearest singularity of $\Psi(\rho)$ and $\Gamma$
describes the asymptotic behaviour of $\Psi(\rho)$ as $\rho\to\infty$, via
\begin{equation}
\Psi(\rho)=2\Gamma
\sqrt\rho
\,K_1\left(
2\sqrt\rho
\right)\left[
1+o(1)
\right],
\end{equation}
where $K_1$ is the modified Bessel function \cite{HMF}. This behaviour
is shown in Appendix \ref{app:a} and $\Gamma$ is evaluated numerically in 
Appendix \ref{app:b}.

Note that the behaviour of $\Phi(x)$ as described in (\ref{eq:phismall}) consists of 2 
different exponential decays going at different rates. As we shall see in the following subsection, something 
very similar occurs if we consider the large-time behaviour of clusters of fixed size, which is not,
however, described by the scaling function.

We now turn to the behaviour of the scaling function $\Phi(x)$ as $x\to\infty$. We obtain
\begin{equation}
\Phi(x)=2\kappa_0^{-1}\exp\left(
-\Lambda x
\right)
\left(
\Lambda x-\frac1{3x}
\right)
\left[
1+o(1)
\right].
\label{eq:largescal}
\end{equation}
In other words, the leading behaviour is exponential decay corrected by a power-law
prefactor $x$, in agreement with the general theory developed in \cite{vDongen4}, which predicts
a prefactor $x^{-\lambda}$, where $\lambda$ is the degree of homogeneity of the reaction kernel. 
We see however a non-trivial correction to this factor, which is a rather unexpected result.

All the above results will be derived in Section \ref{sec:deriv}. 

\subsection{Large clusters}
 In this subsection and the next, we turn to behaviour outside the scaling limit. 
For small times, it is known that solving the Smoluchowski equations
(\ref{eq:basic}) without the loss term leads to the exact small-time 
behaviour, which can be obtained recursively through the {\em Ansatz\/}
\begin{equation}
c_j(t)=\lambda_j t^{j-1}\left[
1+O(t)
\right]
\label{eq:short}
\end{equation}
where the $\lambda_j$ are exactly given by $j2^{-(j-1)}$.

In earlier work \cite{vDongen4,theta}, it had been assumed that both the scaling 
behaviour for $x\gg1$ and the behaviour of large clusters at fixed time would be given by
a qualitatively similar behaviour. We have already seen that the scaling function
for $x\gg1$ behaves differently, see (\ref{eq:largescal}), with a non-trivial subleading correction. In 
Appendix \ref{app:d}, we also show that at a fixed value of time the large clusters 
behave as
\begin{equation}
c_j(t)=j\left(1-\frac2{3j^2R(t)}
\right)R(t)^j\left[
1+o(1)
\right]\qquad(t\to\infty)
\label{eq:largeclust}
\end{equation}
We therefore see that, if we look at subleading behaviour, there is a real difference between 
the recursion describing the behaviour at the smallest times and the behaviour of large clusters,
both when viewed in the scaling limit and when taken at fixed times for $j\to\infty$.

All the above results will be derived in Section \ref{sec:largesizes}.

\subsection{Large-time behaviour of fixed size clusters}
 
Here we ask how clusters of fixed size decay at large times: one finds:
\begin{equation}
\phi_j({s})=\alpha_j\exp(-{s}/j)+\alpha_j^\prime\exp(-4{s}/j).
\label{eq:defalpha}
\end{equation}
Here $\alpha_j$ and $\alpha_j^\prime$ are size-dependent constants
which cannot be evaluated explicitly. Numerical work, however, shows 
that $\alpha_j\simeq j^{-1}$.

We see how the two different 
exponential decays are reflected in the scaling behaviour at $x\ll1$, so
that we may say that the scaling theory is consistent with the large-time 
behaviour for clusters of fixed size. This need not be the case, as is well-known,
for instance, for reaction kernels such as (\ref{eq:exactc}): there one shows that the
scaling theory predicts a large-time decay of $t^{-2}$, for
clusters of size $1\ll j\ll S(t)=t$. Monomers, on the other hand, 
generally have a non-universal decay.

All the above results will be derived in Section \ref{sec:largetimes}.

\section{Scaling behaviour: derivations}
\label{sec:deriv}
Let us first establish a connection between $G(z,{s})$ and $\Phi(x)$. Let us 
take $S(t)$ to be equal to ${s}$. We then have
\begin{eqnarray}
\sum_{j=1}^\infty jc_j(t)\exp\left(-
\frac{j\rho}{{s}}
\right)&=&N({s})\sum_{j=1}^\infty j\phi_j(t)\exp\left(-
\frac{j\rho}{{s}}
\right)\nonumber\\
&=&{s}^2N({s})\,\partial_\rho^2G
\left(
\frac{\rho}{{s}},{s}
\right).
\end{eqnarray}
In the scaling limit, we have,  see (\ref{eq:1.8}):
\begin{equation}
\lim_{{s}\to\infty}\sum_{j=1}^\infty jc_j(t)\exp\left(-
\frac{j\rho}{{s}}
\right)=\int_0^\infty dx\,x\Phi(x)e^{-\rho x}.
\label{eq:defscal}
\end{equation}
We thus have
\begin{equation}
\lim_{{s}\to\infty}
\left[
{s}^2N({s})\,\partial_\rho^2G
\left(
\frac{\rho}{{s}},{s}
\right)
\right]
=\int_0^\infty dx\,x\Phi(x)e^{-\rho x}.
\end{equation}
Now we know that if ${s} G(\rho/{s},{s})$ tends to a limit, that limit is the function 
$\Psi(\rho)$ defined by (\ref{eq:1.10}) and described in greater detail in Appendix \ref{app:a}. 
As we have stated before, it is uniquely determined by the differential equation 
(\ref{eq:1.10}) together with the conditions that it be regular at $x=0$ and positive
and finite for all positive values of $x$. We thus obtain
\begin{equation}
\lim_{{s}\to\infty}
\left[
{s} N({s})
\right]\partial^2_\rho\Psi(\rho )=\int_0^\infty dx\,x\Phi(x)e^{-\rho x},
\label{eq:psiscal2}
\end{equation}
and hence
\begin{equation}
\Psi(\rho )=\left(\lim_{s\to\infty}\left[
s N(s)
\right]
\right)^{-1}
\int_0^\infty dx\frac{\Phi(x)}{x}e^{-\rho x}.
\label{eq:psiscal}
\end{equation}
 This result corresponds to the fundamental equation (\ref{eq:psiscalsumm}) stated in the previous section.
Up to a constant which we shall later show to be equal to $\kappa_0$, 
$\Psi(\rho)$ is the Laplace transform
of $\Phi(x)/x$, where $\Phi(x)$ is the scaling function for the concentrations $c_j(t)$,  again 
as defined by (\ref{eq:1.8}). 
\subsection{Large-time behaviour of moments and connection between $t$ and ${s}$}

Using the  connection between  $\Phi(x)$ and $\Psi(\rho)$ established in (\ref{eq:defscal}) with $\rho=0$, we obtain the identity
\begin{equation}
\int_0^\infty x\Phi(x)dx=1,
\end{equation}
from which we obtain from (\ref{eq:psiscal2}) that
\begin{equation}
\lim_{{s}\to\infty}
\left[
{s} N({s})
\right]=\frac{1}{\Psi^{\prime\prime}(0)}=\kappa_0^{-1},
\label{eq:sns}
\end{equation}
thereby showing 
 (\ref{eq:varmom0}) as well as the claim made after (\ref{eq:psiscal}). Since the derivatives of $\Psi(\rho)$ at the origin 
can all be computed recursively in terms of $\kappa_0$, we have
\begin{eqnarray}
\mu_n({s})&=&\sum_{j=1}^\infty j^nc_j(t)\nonumber\\
&=&N({s}){s}^n\sum_{j=1}^\infty j^n\phi_j({s})
\label{eq:momdef}
\end{eqnarray}
 In the scaling limit, this expression can be expressed in terms of $\partial_\rho^{n+1}\Psi(\rho)$
and yields the result stated in (\ref{eq:mom}).
The asymptotic connection between ${s}$ and $t$  stated in (\ref{eq:taut}) also follows staightforwardly
from (\ref{eq:sns}).

Moreover, along these lines, the fact that $\Psi(0)=1$ tells us that
\begin{equation}
\lim_{{s}\to\infty}\left(
{s}^2\sum_{j=1}^\infty 
\frac{c_j}{j}
\right)=\kappa_0^{-1},
\end{equation}
which leads to the result in (\ref{eq:negmoma}).
Finally, the result obtained in Lemma \ref{lemma:integ} of Appendix \ref{app:a} yields a result for the asymptotic
behaviour of yet another moment, namely $\mu_{-2}$, as obtained in (\ref{eq:negmomb}).
Thus the asymptotic behaviour of all moments with $n\geq-2$ is determined in elementary terms
from the knowledge of $\kappa_0$.

Even though it is possible to obtain the $m_n^{(\infty)}$ explicitly by a recursion, it is
not possible to obtain an explicit expression for them. Determining their symptotic 
behaviour as $n\to\infty$ is therefore not trivial. Clearly, this depends on the 
nature of the singularity of $\Psi(\rho)$ closest to the origin. 
From the results of Appendix \ref{app:a}, Lemma \ref{lemma:anal},
we see that the coefficients of the series development of $\Psi(\rho)$ are 
real and of alternating sign, so that this singularity, which we denote by
$-\Lambda$, must lie on 
the negative real axis. Its value cannot be expressed in elementary
terms from $\kappa_0$, but can be computed numerically by 
integration of (\ref{eq:1.10}) up to values of $\rho$ close to 
$-\Lambda$. This is carried out in detail in Appendix \ref{app:b}, together with other numerical evaluations.
Its position is found to have the value
$\Lambda\approx1.576\,132\ldots$
 
 The leading behaviour of the singularity at $-\Lambda$ is readily found, by
 matching orders of divergence, to be a simple pole, with a residue 
 $2\Lambda$. In other words
 \begin{equation}
\Psi(\rho)=\frac{2\Lambda}{\rho+\Lambda}\left[
1+o(1)
\right]
\end{equation}
for $\rho$ near $-\Lambda$, up to singular terms of higher order. Deciding whether or
not such corrections exist is a bit more intricate, and is carried out in Appendix \ref{app:c}. One finds
that there is in fact a correction given by
 \begin{equation}
\Psi(\rho)=\left[
\frac{2\Lambda}{\rho+\Lambda}-2-\frac23
\left(
\rho+\Lambda
\right)\ln\left(
\rho+\Lambda
\right)
\right]\left[
1+o(1)
\right].
\label{eq:singdet}
\end{equation}
From this then follows the result  claimed in (\ref{eq:momasym}), where
the correction in $n^{-2}$ corresponds to the singularity found in (\ref{eq:singdet}). 

In all the preceding considerations, the scaling limit has been used. It may be asked whether this is legitimate.
The problem is that small clusters, even in the infinite time limit, are not necessarily described by 
the scaling limit, as discussed for instance in \cite{mob03}. The results above thus strictly speaking do not apply to the moments as defined by (\ref{eq:momdef}),
but rather to moments defined as
\begin{equation}
\tilde{\mu}_{n,\epsilon}(s)=\sum_{j\geq\epsilon s}j^nc_j(s).
\label{eq:momscaldef}
\end{equation}
The basic result (\ref{eq:moma}) should thus read rather
\begin{equation}
m_n^{(\infty)}=\lim_{\epsilon\to0}\lim_{s\to\infty}\left[
s^{-(n-1)}\tilde{\mu}_{n,\epsilon}(s)
\right].
\label{eq:momscal}
\end{equation}
It turns out, however, that these  distinctions are unnecessary, and that (\ref{eq:moma})
is correct as it stands. The proof is a bit intricate and is thus relegated to Appendix \ref{app:e}. 

\subsection{Behaviour of small clusters}

Another important issue is the behaviour of $\Phi(x)$ close to the origin, that is, 
the behaviour of cluster of size much less than the typical size $S(t)\propto s\propto\sqrt t$. This corresponds to 
the behaviour of $\Psi(\rho)$ for $\rho\to\infty$. As shown in Appendix \ref{app:a}, Lemma \ref{lemma:dec},
$\Psi(\rho)$ decays as $2\Gamma\sqrt\rho\,K_1(2\sqrt\rho)$, where $\Gamma$ is an undetermined 
positive constant and $K_1$ is a modified Bessel function \cite{HMF}. Asymptotically, this means
that
\begin{equation}
\Psi(\rho)=\sqrt\pi\,\Gamma\rho^{1/4}\exp\left(
-2\sqrt\rho
\right)\left[
1+o(1)
\right]
\end{equation}
Note again that the constant $\Gamma$
does not have an explicit expression in terms of $\kappa_0$, but it can be obtained to high accuracy by numerical integration
of (\ref{eq:1.10}), which leads to the value $\Gamma\simeq1.707\,87$. The details are discussed in Appendix \ref{app:b}. 

It is readily calculated that
\begin{equation}
2\sqrt\rho \,K_1(2\sqrt\rho)=\int_0^\infty \frac{\exp(-1/x)}{x^2}e^{-\rho x}\,dx,
\end{equation}
which leads to  the leading part of the result stated in (\ref{eq:phismall}).
This is in   excellent agreement with the results of \cite{mim09} as well as with the 
general scaling results derived for instance in \cite{ley03}. 

 To obtain the rest of the result stated in (\ref{eq:phismall}) we ask about the next-to-leading 
asymptotic small-$x$ behaviour of $\Phi(x)$, or
correspondingly, the next-to-leading asymptotic large-$\rho$ behaviour of $\Psi(\rho)$. This 
can be obtained as follows: define
\begin{equation}
H(\rho)=2\Gamma\sqrt\rho\,K_1(2\sqrt\rho),
\end{equation}
which is the exact asymptotic behaviour of $\Psi(\rho)$ and consider
\begin{equation}
\Psi_1(\rho)=\frac{\Psi(\rho)}{H(\rho)}.
\end{equation}
Clearly $\Psi_1(\rho)$ approaches 1 as $\rho\to\infty$. If we now rewrite (\ref{eq:1.10}) for $\Psi_1(\rho)$,
we find
\begin{equation}
\rho\Psi_1^{\prime\prime}(\rho)+2\rho\frac{H^\prime(\rho)}{H(\rho)}
\Psi_1^\prime(\rho)=H(\rho)\Psi_1(\rho)\Psi_1^\prime(\rho)+2H^\prime(\rho)\Psi_1(\rho)^2.
\label{eq:nextorder}
\end{equation}
Replacing $\Psi_1$ by its limiting value we obtain
\begin{equation}
\rho\Psi_1^{\prime\prime}(\rho)+2\rho\frac{H^\prime(\rho)}{H(\rho)}
\Psi_1^\prime(\rho)=H(\rho)\Psi_1^\prime(\rho)+2H^\prime(\rho).
\label{eq:nextorder1}
\end{equation}
This is a first order linear equation for $\Psi_1^\prime(\rho)$. We now perform the substitution
\begin{equation}
\Psi_1^\prime(\rho)=H(\rho)^{-2}\chi(\rho)
\end{equation}
which leads to
\begin{equation}
\rho\chi^\prime(\rho)=H(\rho)\chi(\rho)+H^\prime(\rho) H(\rho)^2.
\end{equation}
Since $H(\rho)$ is rapidly decaying  as $\rho\to\infty$ , we see that 
$\chi(\rho)$ tends to a limiting value, which is of the
order of the integral of the inhomogeneity, that is, $H(\rho)^3$ as $\rho\to\infty$.
We thus conclude that
\begin{equation}
\Psi_1^\prime(\rho)\approx H(\rho)^2\chi(\rho)\approx H(\rho)\qquad(\rho\to\infty). 
\end{equation}
From this follows that the order of magnitude of the correction to scaling is $\rho H(\rho)^2$, in other words
\begin{equation}
\Psi_1(\rho)=1+const.\cdot \rho^{3/2}\exp\left(
-4\sqrt\rho
\right)
\end{equation}
But the Laplace transform of $x^{-9/2}\exp(-4/x)$ is given by
\begin{equation}
\frac{\sqrt\pi }{1024} e^{-4 \sqrt{\rho
   }} \left(64 \rho ^{3/2}+96 \rho
   +60 \sqrt{\rho }+15\right)
\end{equation}
which has the same large-$\rho$ asymptotic behaviour as $\Psi_1(\rho)$, implying
that $\Phi(x)/x$ has the same subdominant small-$x$ behaviour as $x^{-9/2}\exp(-4/x)$. 
In other words, the correction goes to zero as $x\to0$ exponentially faster than the
leading behaviour, and with a different correction exponent. Explicitly this 
yields  the full expression given in (\ref{eq:phismall}). 

\subsection{Behaviour of large clusters}
\label{subsec:largescal}

We may also ask how $\Phi(x)$ behaves as $x\to\infty$, in other words, 
how does the concentration of large clusters behave. Since $\Psi(\rho)$
is well-defined and finite over a finite range of negative values of $\rho$,
it follows from (\ref{eq:psiscal}) that $\Phi(x)/x$ decays exponentially as $x\to\infty$.
More information is obtained by referring to our results on the nearest singularity
of $\Psi(\rho)$, which is at $\rho=-\Lambda$. As discussed earlier and shown in Appendix \ref{app:c}, 
see (\ref{eq:singdet}),
this singularity is a simple pole with residue $2\Lambda$ and a correction term 
of the form $(\rho+\Lambda)\ln(\rho+\Lambda)$. 
We therefore have:
\begin{equation}
\kappa_0\int_0^\infty\frac{\Phi(x)}{x}e^{-\rho x}\,dx=
\left[
\frac{2\Lambda}{\rho+\Lambda}
-2-\frac23
\left(
\rho+\Lambda\right)
\ln
\left(
\rho+\Lambda
\right)
\right]
\left[
1+o(1)
\right]
\end{equation}
in the limit $\rho\to-\Lambda$
Since the Laplace transform $\chi(\rho)$ of 
\begin{equation}
-\frac23\frac{\exp\left(
-\Lambda x
\right)}{x^2+b}
\end{equation}
has the same asymptotic behaviour for $\rho\to-\Lambda$ as the correction 
to $\Psi(\rho)$ for any positive value of $b$, we obtain
 for $\Phi(x)$ as $x\to\infty$ the asymptotic expression stated earlier in (\ref{eq:largescal}).
The approximate value of $\Lambda$ is computed in Appendix \ref{app:b}. 

\section{Behaviour at large sizes for fixed times}
\label{sec:largesizes}
The behaviour at large sizes for fixed times has been assumed to be similar 
to the behaviour for large sizes at very small times. This can be understood 
qualitatively by imagining that, at time $t$, the system has acquired a typical 
size $S(t)$. We may now coarse grain the system in such a way that 
all aggregates within a (large) multiple of $S(t)$ are viewed as momomers.
On that scale, all aggregates of large size are much larger than all aggregates
that have been produced at that time, and we may therefore proceed similarly
 to the case in which monomers altogether dominate the cluster size distribution. 

In the limit of small times, we may make the  {\em Ansatz} (\ref{eq:short}), which solves 
the system (\ref{eq:basic}) in leading order of $t$ for small $t$. 
Putting (\ref{eq:short}) into (\ref{eq:basic}) leads to the recursion
\begin{subequations}
\begin{eqnarray}
\lambda_1&=&1
\label{eq:reclargea}
\\
(j-1)\lambda_j&=&\frac12\sum_{k=1}^{j-1}K(k,j-k)\lambda_k\lambda_{j-k}.
\label{eq:reclargeb}
\end{eqnarray}
\label{eq:reclarge}
\end{subequations}
where (\ref{eq:reclargea}) follows from the assumption
that the initial distribution has concentration $1$ of monomers.
This simply reflects the fact that at short times, loss terms are
dominated by the (recursive) production terms.

 Putting the expression
\begin{equation}
\lambda_j=2^{-(j-1)}j,
\label{eq:beta}
\end{equation}
into (\ref{eq:reclarge}), we find that it provides the unique solution in the case we study.
Again this is in good agreement with the general predictions of \cite{vDongen4,theta}. Indeed,
in these references it is shown that the behaviour of the general recursion (\ref{eq:reclarge})
is given by
\begin{equation}
\lambda_j\simeq j^{-\lambda} R^j,
\end{equation}
where $R$ is a non-universal constant (except when $\nu=1$, which is a singular case). Since we have 
$\lambda=\mu=-1$, the result (\ref{eq:beta}) is in full agreement with the predictions. 
Again, these results were also derived in \cite{mim09}. 

Note that the behaviour at finite times is in agreement with the scaling prediction to leading order.
To subleading order, however, we have seen in (\ref{eq:largescal}), that the singularity of $\Psi(\rho)$
is a pole modified by a correction of order $(\rho+\Lambda)\ln(\rho+\Lambda)$. This implies a correction
of relative order $1/x^2$ to the leading behaviour $xe^{-x}$ of the scaling function for large $x$, see
(\ref{eq:largescal}). 
This implies that there is a discrepancy between the small-time approximation, which leads to an 
exact exponential decay for the concentrations at large sizes, and the scaling function, which 
displays a correction $-2/(3x^2)$.

In Appendix \ref{app:d} we present arguments suggesting that in fact there are  in fact also
similar corrections to the simple
pole singularity of $G(z,t)$ at fixed finite $t$, contrary to the small-time approximation.
This means that both the large-size limit of the aggregate size distribution at fixed times
and the scaling limit for large values of $x$ behave similarly. On the other hand, the simpler
approximation involving a recursion relation valid for small times does not accurately capture these features. 

\section{Behaviour at large times for aggregates of fixed size}
\label{sec:largetimes}
In the previous Section we have analysed the behaviour of aggregates having a size
proportional to the typical size, namely of size $xS(t)$, for values of $x$ which are either $x\ll1$
or $x\gg1$. In the following, we shall look at aggregates of fixed size, in the limit ${s}\to\infty$.

Consider the system of equations (\ref{eq:1.5}). Inductively it is easily seen that 
each $\phi_j({s})$ is a finite linear combination of decaying exponentials. Indeed, let
\begin{equation}
\phi_j^\prime({s})=F_j({s})-\phi_j({s})/j
\end{equation}
and assume $F_j({s})$ to consist of a finite linear combination of decaying exponentials. 
It is then easy to show that $\phi_j({s})$ is a linear combination of these same exponentials and 
$\exp(-{s}/j)$ for instance using the Laplace transform approach. 

If we denote
by $\sigma_{j,k}$ the decay rates and by $\Sigma_j$ the set of all the rates that correspond
to $\phi_j({s})$, we have
\begin{equation}
\phi_j({s})=\sum_k \alpha_k\exp(-\sigma_{j,k}{s}).
\end{equation}
Inductively we also show that
\begin{equation}
\Sigma_j=\bigcup_{k=1}^{j-1}\left(
\Sigma_k+\Sigma_{j-k}
\right)\cup\left\{
\frac1j
\right\}
\end{equation}
where the sum of two sets is defined as the set of all possible sums between elements 
of both sets. 

\begin{figure}
\includegraphics[scale=0.8]{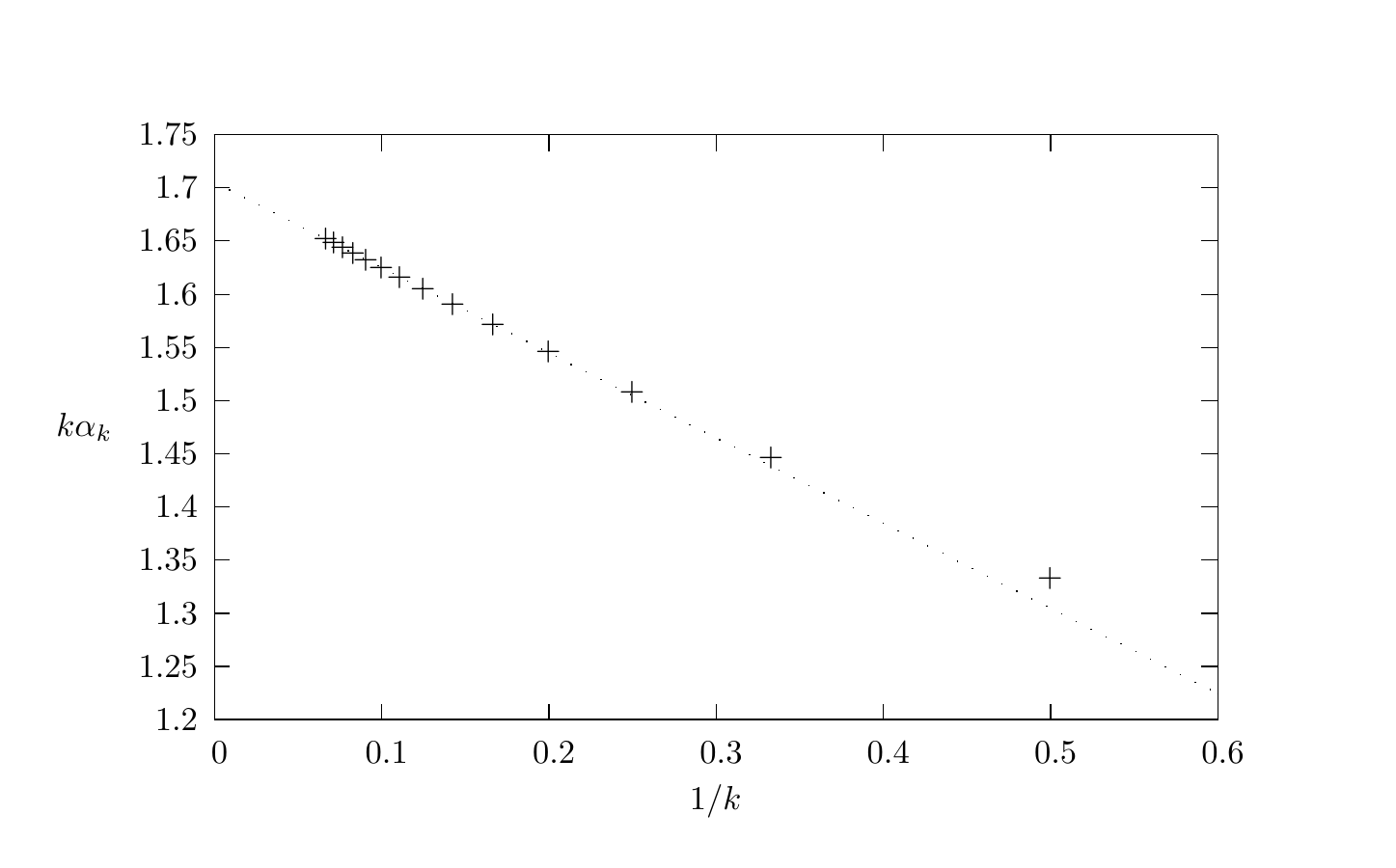}
\caption{
Value of $k\alpha_k$ for $2\leq k\leq15$ plotted as a function of $1/k$. 
One sees that the extrapolation to a limiting value seems reasonable. The line is a least-squares fit through
the points with $k\geq5$ and yields an asymptotic value of $1.71$. 
}
\label{fig:5}
\end{figure}

From this follows that the slowest decay rate in $\Sigma_j$ is always $1/j$. The next
lowest decay rate in $\Sigma_j$ arises from the sum of the 2 lowest rates of 
$\Sigma_{j/2}$. This means that the next lowest decay rate of $\Sigma_j$
is of the order $4/j$.

This implies that each $\phi_j({s})$ has exactly one exponential contribution of 
the form $\exp(-{s}/j)$. In other words, the large-time decay of $\phi_j({s})$
is exactly $\alpha_j\exp(-{s}/j)$ for a given value of $\alpha_j$. Hence the 
behaviour of $c_j({s})$ is given by
\begin{equation}
c_j({s})=\frac{\alpha_j}{{s}}e^{-{s}/j}\qquad({s}\to\infty)
\end{equation}
This is qualitatively similar to the scaling result, which states that for $x\ll1$
\begin{equation}
\Phi(x)=\frac{\Gamma}{\kappa_0}\frac{\exp(-1/x)}{x}\left[
1+o(1)
\right]
\end{equation}
and this would suggest that
\begin{equation}
\alpha_j\simeq j^{-1}.
\label{eq:alpha}
\end{equation}
It is possible, of course, to evaluate the $\alpha_j$ recursively, but it is not possible
to find for them an explicit expression. On the other hand, one can evaluate the $\alpha_j$
explicitly for $1\leq j\leq15$ and verify that, at least for these values, (\ref{eq:alpha})
appears to hold to a good approximation: we show the product $k\alpha_k$ in Figure \ref{fig:5}
for $2\leq k\leq15$, as a function of $1/k$ and it appears reasonably to extrapolate to a 
well-defined value. The explicit values of $\alpha_j$ are also tabulated in Table \ref{tab:1}. 

We thus find the large-time behaviour of fixed size aggregates to be consistent with the scaling behaviour
of aggregates of size small with respect to the typical size. While such agreement is not in itself surprising,
it should be pointed out that this is not a necessary feature of aggregating models: several counterexamples
have been discussed, for instance, in \cite{ley03}. Indeed, this coincidence goes even beyond the leading order:
as noted above, the next to leading order for the large-time exponential decay consists of decay rates
of the order of $4/j$, so that one has the result stated in (\ref{eq:defalpha}). 

Here the $\alpha_j^\prime$ are the prefactors corresponding to the decay rate $4/j$. Again this 
fits well with the behaviour of the scaling function, which displays both a decay of type $\exp(-1/x)$,
with a correction of order of $\exp(-4/x)$. On the other hand, the order of the $\alpha_j^\prime$
cannot be estimated, as there are too few expressions for $\phi_j({s})$ available. 

\begin{table}
 \caption{\label{tab:1}
 The values of $\alpha_j$ as defined in (\ref{eq:alpha}) for $2\leq j\leq15$, evaluated from exactly
computed fractions.
 }
 \begin{ruledtabular}
 \begin{tabular}{cc}
2&0.666666666666667\\   
3&0.482142857142857\\   
4&0.376828110161443\\   
5&0.309065279025988\\   
6&0.261879521457270\\   
7& 0.227157280354022\\   
8&0.200546422642681\\   
9&0.179506571151271\\   
10&0.162456450838603\\   
11&0.148360805859407\\   
12&0.136513680323195\\   
13&0.126417243694629\\   
14&0.117710416149960\\   
15&0.110124961934082\\   
 \end{tabular}
 \end{ruledtabular}
 \end{table}

\section{Conclusions}
\label{sec:conclusions}
To summarise, we have derived an exact ordinary differential of second order for the Laplace transform of the 
scaling function of the solution for Smoluchowski's equations (\ref{eq:basic}) for the rate kernel
\begin{equation}
K(k,l)=\frac1{k}+\frac1{l}.
\end{equation}
We show that this differential equation has a unique solution with the properties required of the Laplace transform 
of a scaling function. From this we obtain a large number of accurate asymptotic results involving the 
detailed behaviour of the scaling function both for $x\gg1$ and $x\ll1$. Similarly we obtain 
the amplitudes for the large time behaviour of all the integer moments 
\begin{equation}
\mu_n({s})=\sum_{j=1}^\infty j^nc_j({s})
\end{equation}
for $n\geq-2$, as well as the detailed asymptotic connection between ${s}$ and $t$. 
The detailed behaviour of the scaling function coincides quite well with the general predictions of the scaling theory,
as described for instance in \cite{ley03}.  Finally we also obtain results for behaviour which in 
principle is not accessible to scaling, such as the behaviour of clusters at large sizes and fixed time, or else
the behaviour of clusters of fixed size at large times. 

As stated in Section \ref{sec:notation}, the convergence of the exact solution to a scaling limit follows from the work of Norris
\cite{norris}. Whether a more precise description of this convergence could be obtained by a detailed study of the
solution of (\ref{eq:1.7}) is left for future work.

\begin{acknowledgments}
The author gratefully acknowledges financial support
of the grants UNAM--DGAPA PAPIIT IN113620
and CONACyT 254515.
\end{acknowledgments}
\appendix

\section{Qualitative behaviour of the fundamental equation}
\label{app:a}
\setcounter{lemma}{0}
In the following, we shall prove the following theorem
\begin{theorem}

The solutions of the problem
\begin{equation}
\rho \Psi^{\prime\prime}=\Psi\left(
\Psi^\prime+1
\right)\label{eq:a1}
\end{equation}
which are {\em smooth} and {\em positive} at $\rho =0$ and satisfy 
$\Psi^{\prime\prime}(0)\neq0$ fall into three mutually disjoint classes:
\begin{enumerate}
\item Those which cross the $\rho $ axis at some value $\rho _0$ of $\rho $. These
then remain negative for some values $\rho >\rho _0$.

\item Those which reach a positive minimum at some positive value $\rho _0$ of $\rho $. 
These grow for all $\rho >\rho _0$, until they eventually diverge. In any case, 
they do not tend to zero as $\rho\to\infty$

\item a unique function with $\Psi( 0)=1$, $\Psi^\prime(0)=-1$ and 
$\Psi^{\prime\prime}(0)=\kappa_0$
given by a unique positive value satisfying
$1.45582<\kappa_0<1.45583$. This function 
goes to zero as $const.\cdot2\sqrt \rho  K_1(2\sqrt \rho )$ as $\rho \to\infty$ 
on the positive real axis. 
\end{enumerate}
\end{theorem}

The results follow from a tedious sequence of lemmas
\begin{lemma}
Any solution of (\ref{eq:a1}) {\em smooth} at $\rho =0$ and with $\Psi( 0)>0$ and 
$\Psi^{\prime\prime}(0)\neq0$
satisfies $\Psi( 0)=-\Psi^\prime(0)=1$
\end{lemma}
Indeed, if $\Psi^\prime(0)\neq-1$, then, since $\Psi( 0)\neq0$, $\Psi^{\prime\prime}(\rho)$
diverges to $\infty$ as $\rho\to0$, so $\Psi( \rho)$ cannot be smooth. One now rewrites 
(\ref{eq:a1}) as follows
\begin{equation}
\rho (1-\Psi)\Psi^{\prime\prime}=\Psi\left(
\Psi^\prime+1-\rho \Psi^{\prime\prime}
\right).
\label{eq:a2}
\end{equation}
Since we have assumed the smoothness of $\Psi( \rho )$ near $\rho =0$, by Taylor's
theorem applied to $\Psi^\prime(\rho )$ for $\rho $ near $0$, 
the r.h.s.\ of (\ref{eq:a2}) is of order $O(\rho ^2)$ as $\rho \to0$. Since 
$\Psi^{\prime\prime}(0)\neq0$, it follows that the l.h.s.\ can only be of the same  order
if $1-\Psi$ vanishes linearly in $\rho $ as $\rho \to0$. We have thus shown the lemma. Note that 
the hypothesis $\Psi^{\prime\prime}(0)\neq0$ is indeed necessary, since
$\Psi=b-\rho $ is an exact solution of (\ref{eq:a1}) for all $b$.
 Note that we exclude the case in which $\Psi^{\prime\prime}(0)=0$ because
these solutions do not satisfy the requirements of the problem at hand: indeed 
$\Psi(\rho)$ is the Laplace transform
of the scaling function $\Phi(x)/x$, so that the second derivative of $\Psi(\rho)$ corresponds to the first moment of 
the positive function $\Phi(x)$, which cannot vanish.

In the following, we shall exclusively limit ourselves to solutions 
satisfying the hypotheses of Lemma 1. We shall denote them as regular solutions. 

\begin{lemma}
There is a family of regular solutions of (\ref{eq:a1}) indexed by
the arbitrary real parameter $\kappa=\Psi^{\prime\prime}(0)$. These solutions 
are analytic in an appropriately small neighbourhood of the origin.
\label{lemma:anal}
\end{lemma}

Define $z=\Psi-1+\rho $. (\ref{eq:a1}) then becomes
\begin{equation}
\rho z^{\prime\prime}-z^\prime=zz^\prime-\rho z^\prime.
\label{eq:a2.1}
\end{equation}
From Lemma 1, we know that $z(0)=z^\prime(0)=0$. Now if we consider
the leading order behaviour of both terms of (\ref{eq:a2.1}) if $z=\kappa \rho ^2/2$,
we find that the equation is identically satisfied for all values of $\kappa$.
If we now substitute
\begin{equation}
z(\rho )=\frac{\kappa \rho ^2}{2}+\sum_{n=3}^\infty (-1)^na_n\rho ^n
\label{eq:a2.2}
\end{equation}
formally in (\ref{eq:a2.1}), we obtain the following recurrence relations
\begin{subequations}
\begin{eqnarray}
a_2&=&\frac\kappa2.\label{eq:a2.3a}\\
a_{m+1}&=&\frac{1}{m^2-1}\left(
\sum_{k=2}^{m-1}ka_ka_{m-k+1}+ma_m
\right)
\qquad(m\geq2)
\label{eq:a2.3b}
\end{eqnarray}
\label{eq:a2.3}
\end{subequations}
To show that the series defined by (\ref{eq:a2.2}) has a finite radius of convergence,
we choose $R>\max(a_2,1)$ arbitrary. It is then straightforward to show that 
\begin{equation}
a_m\leq R^m,
\end{equation}
so that the series converges in a circle of radius $R$. 
The Lemma is thus completely proved.

\begin{lemma}
Let $\Psi( \rho )$ be any regular solution of (\ref{eq:a1}) with the property of being integrable 
over the positive real axis. Then one has
\begin{equation}
\int_0^\infty \Psi( \rho )\,d\rho =\frac32.
\label{eq:a3}
\end{equation}
\label{lemma:integ}
\end{lemma}
Note that we say nothing here concerning either the existence or
the  uniqueness of such solutions.
To prove (\ref{eq:a3}), we rewrite (\ref{eq:a1}) as
\begin{equation}
\frac{d^2}{d\rho^2}\left[
\rho\Psi(\rho)
\right]=\frac{d}{d\rho }\left(
\frac{\Psi(\rho)^2}{2}+2\Psi(\rho)
\right)+\Psi(\rho)
\label{eq:a4}
\end{equation}
and integrate from $0$ to infinity on both sides, from which the lemma immediately 
follows from the fact that $\Psi( 0)=1$ and $\Psi^\prime(0)=-1$.

\begin{lemma}
Let $\Psi(\rho)$ be a regular solution of (\ref{eq:a1}). Then $\Psi^{\prime\prime}(\rho)>0$
as long as $\Psi(\rho)>0$.
\end{lemma}

This follows from the following remark
\begin{equation}
\Psi^{\prime\prime}(\rho)=\Psi(\rho)\frac{\Psi^\prime(\rho)-\Psi^\prime(0)}{\rho}
=\Psi(\rho)\Psi^{\prime\prime}(\xi)
\label{eq:arec}
\end{equation}
where $0\leq\xi\leq\rho$, which follows from the Intermediate Value Theorem. The Lemma follows,
since, if $\Psi^{\prime\prime}(\rho)$ changed sign, (\ref{eq:arec}) would be violated at the first
point where $\Psi^{\prime\prime}(\rho)=0$.

\begin{lemma}
Let $\Psi( \rho )$ be a regular solution of (\ref{eq:a1}) and assume that  there is an
$\rho _0>0$ such that $\Psi^\prime(\rho _0)=0$ and $\Psi( \rho _0)>0$. 
Then $\Psi(\rho)$ grows monotonically for $\rho\geq\rho_0$. 
\end{lemma}

This follows from the preceding lemma. In fact, it can be shown 
that in this case there is a $\rho_1>\rho_0$ where $\Psi(\rho)$ diverges. However, it is not necessary to
show this: the Lemma's conclusion shows that such a $\Psi(\rho)$ is unacceptable, since it
does not tend to zero as $\rho\to\infty$.

\begin{lemma}
Let $\Psi( \rho )$ be a regular solution of (\ref{eq:a1}) such that $\Psi( \rho _0)=0$ for some
$\rho _0>0$. Then $\Psi( \rho )$ is negative for some $\rho >\rho _0$.
\end{lemma}

Let $\rho_0$ be the smallest $\rho$ such that $\Psi(\rho_0)=0$. Thus $\Psi^\prime(\rho_0)\leq0$. If the inequality is strict, 
the Lemma follows by Taylor's theorem. But $\Psi^\prime(0)=0$ violates the uniqueness theorem for ODE's, since 
$\Psi(\rho)=0$ is a solution of (\ref{eq:a1}). The Lemma follows.

\begin{lemma}
Let $\Psi( \rho )$ be a regular solution of (\ref{eq:a1}), with $\Psi( \rho )>0$ for all
$\rho >0$ and tends to zero as $\rho \to\infty$. Then there is a constant $\Gamma$
such that 
\begin{equation}
\lim_{\rho \to\infty}\frac{\Psi( \rho )}{2\sqrt \rho\,K_1(2\sqrt{\rho} )}=\Gamma
\label{eq:a5.1}
\end{equation}
\label{lemma:dec}
\end{lemma}

\begin{figure}
\includegraphics[scale=0.8]{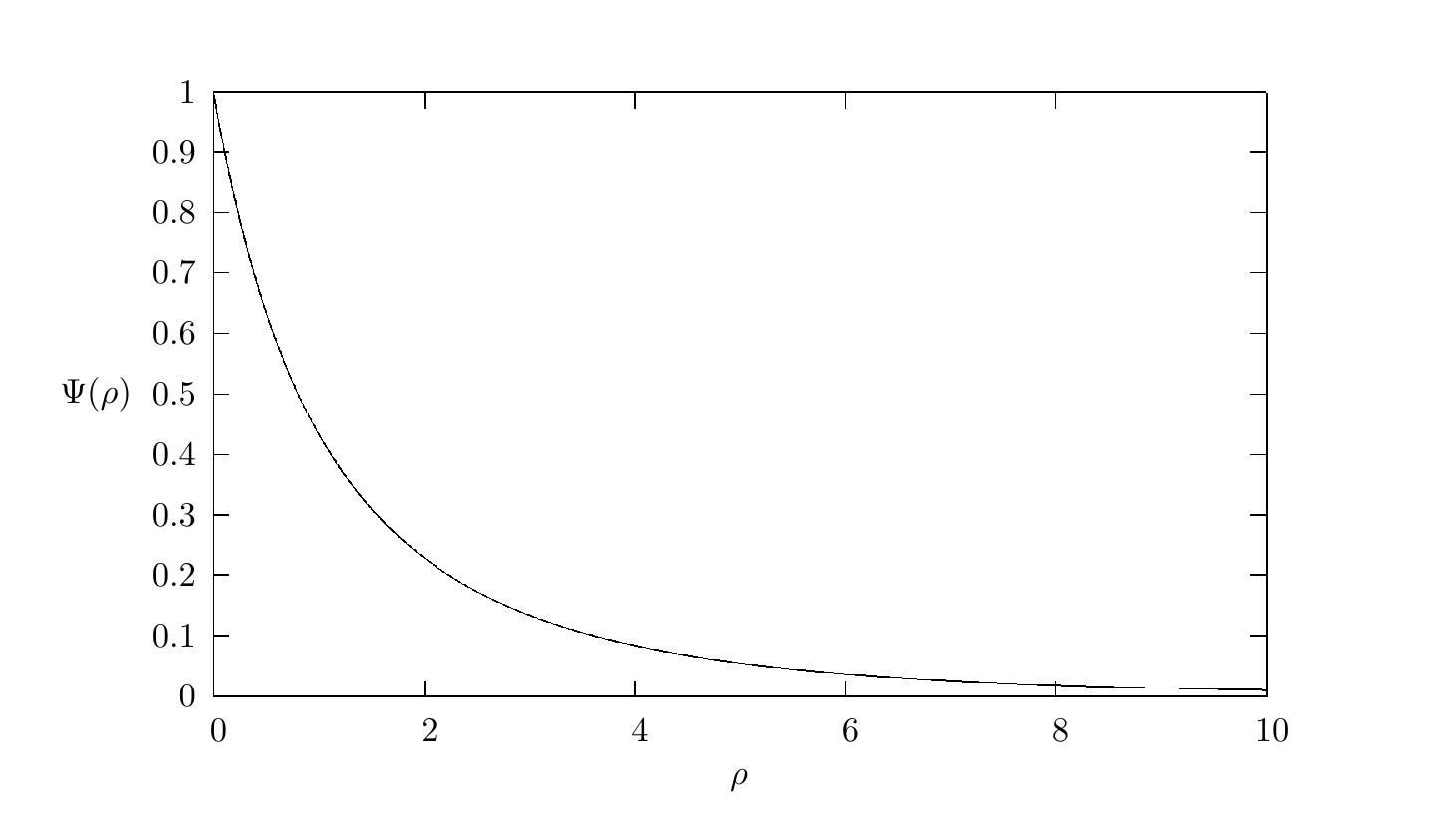}
\includegraphics[scale=0.8]{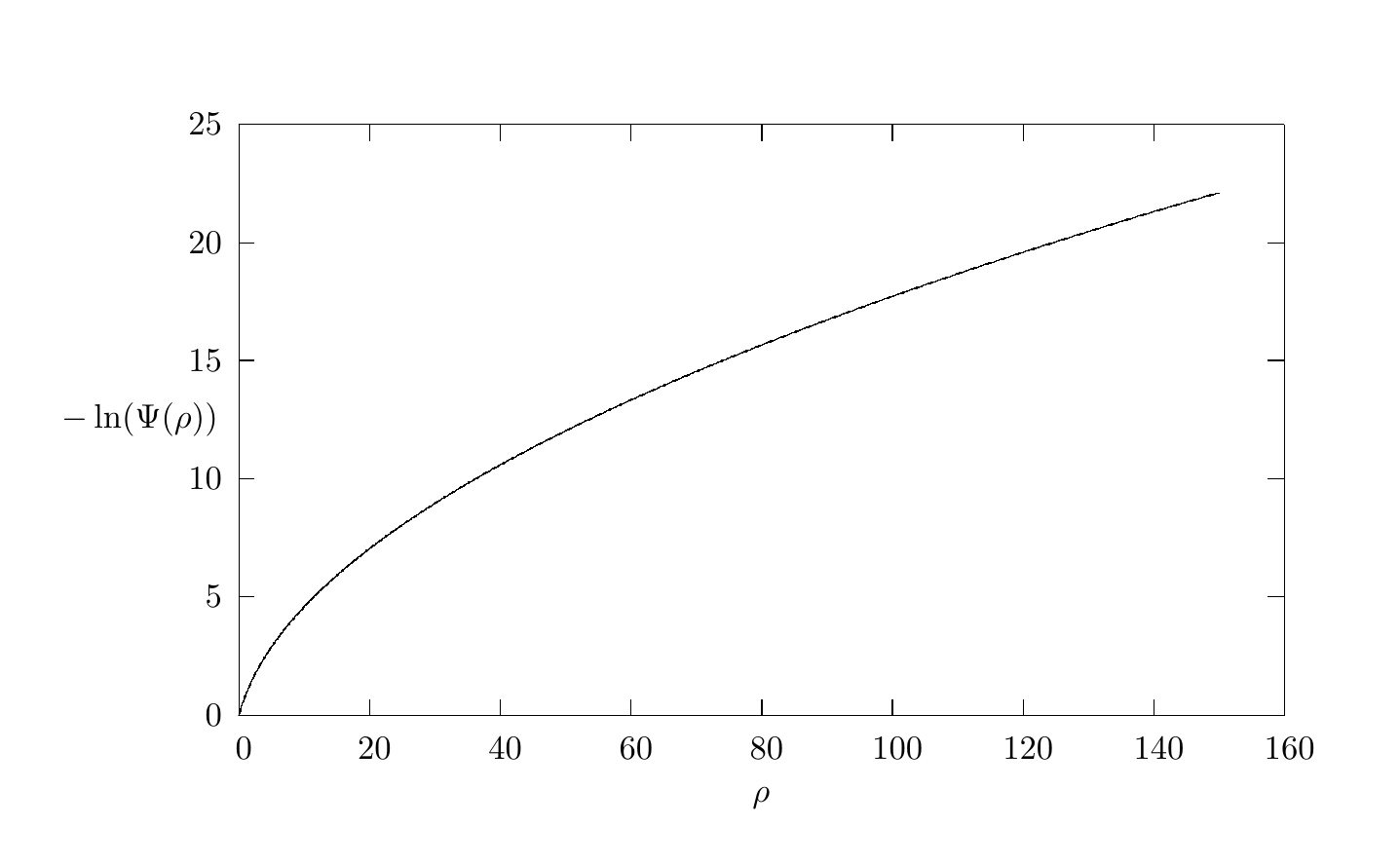}
\caption{
Plot of $\Psi(\rho)$ both on a direct and on a logarithmic scale. The 
strongly non-exponential decay is clearly noticeable.
}
\label{fig:6}
\end{figure}

From Lemma 5 we see that such a solution may neither have a minimum, nor ever 
become zero. It thus decays  monotonically to zero. Since it follows from (\ref{eq:arec})
that $\Psi^{\prime\prime}(\rho )>0$ for all $\rho >0$, $\Psi^\prime(\rho )$ is monotonically
increasing for all $\rho >0$. It follows that $\Psi^\prime(\rho )$
also tends to zero as $\rho \to\infty$. Under these conditions, it is clear that 
(\ref{eq:a1}) can, for $\rho \gg1$, be approximated by the linear equation
\begin{equation}
\rho \Psi^{\prime\prime}=\Psi
\label{eq:a6}
\end{equation}
which has the two solutions $\sqrt \rho\, I_1(2\sqrt\rho)$ and $\sqrt\rho \,K_1(2\sqrt\rho )$. 
The Lemma is proved, since the former diverges exponentially. 

\begin{lemma}
Let $\Psi_1(\rho )$ and $\Psi_2(\rho )$ be  the two solutions such that $\Psi_{i}^{\prime\prime}(0)
=\kappa_i$ for $i=1,2$ and let $\kappa_1<\kappa_2$. Then for all $\rho >0$ such that 
no $\Psi_i(\rho )$ has either diverged or become negative, $\Psi_1(\rho )<\Psi_2(\rho )$. 
\end{lemma}

From the Taylor series of $\Psi_i(\rho )$, it immediately follows that $\Psi_1(\rho )<\Psi_2(\rho )$
as well as $\Psi_1^\prime(\rho )<\Psi_2^\prime(\rho )$
on a sufficiently small open interval $(0, \epsilon)$.  Assume now  that the
Lemma's conclusion fails. Then there is a smallest
$\rho _0$ for which $0<\Psi_1(\rho _0)=\Psi_2(\rho _0)<\infty$. Again, since the two solutions
are not identical, it follows from the uniqueness theorem for ODE's that 
$\Psi_1^\prime(\rho _0)\neq \Psi_2^\prime(\rho _0)$ and thus that 
$\Psi_1^\prime(\rho _0)>\Psi_2^\prime(\rho _0)$. There hence exists a smallest
 $\rho _1$ with $0<\rho _1<\rho _0$, such that $\Psi_1^\prime(\rho _1)=\Psi_2^\prime(\rho _1)$. 
 Since $\Psi_i^{\prime\prime}(\rho )>0$ for all $0\leq \rho \leq \rho _1$, $\Psi_i^\prime(\rho )$
 is growing in this interval.  Since $\Psi_1^\prime(\rho )<\Psi_2^\prime(\rho )$ for $0<\rho <\rho _1$,
 we have $\Psi_1^{\prime\prime}(\rho _1)>\Psi_2^{\prime\prime}(\rho _1)$. But this, together
 with the already established facts that $\Psi_1(\rho _1)<\Psi_2(\rho _1)$ (since $0<\rho _1<\rho _0$)
 and that $\Psi_1^\prime(\rho _1)=\Psi_2^\prime(\rho _1)$ (by definition of $\rho _1$) lead to a 
 contradiction with (\ref{eq:a1}). The Lemma is thus proved. 

\begin{lemma}
There exist solutions that go negative, and solutions with
a positive minimum. 
\end{lemma}
If we choose $\kappa=0$, then the corresponding solution is $\Psi( \rho )=1-\rho $.
This becomes negative at $\rho =1$. By continuous dependence
on initial conditions, solutions with sufficiently small positive
values of $\kappa$ will go negative at some positive value of $\rho $. 
It remains to show that, for $\kappa$ sufficiently large, $\Psi(\rho)$ has 
a minimum.  Using the Taylor series with remainder one obtains,
using $\Psi^{\prime\prime\prime}(0)=-\kappa/3$: 
\begin{equation}
\Psi^\prime(\rho)=-1+\kappa\rho-\kappa\xi^2
\end{equation}
with $0\leq\xi\leq\rho$. From this follows
\begin{equation}
\Psi^\prime\left(
\frac2\kappa
\right)=1+O(\kappa^{-1})
\end{equation}
which is positive for sufficiently large $\kappa$. The lemma follows since $\Psi^\prime(0)=-1$. 

\begin{lemma}
There exists a unique solution satisfying the hypotheses of Lemma 7.
\end{lemma}

Using the theorem concerning the continuous dependence of the solutions of 
ODE's on initial conditions, we see that the set $S_1$ of $\kappa$ such that 
the solution goes negative is open. So is the set $S_2$
of $\kappa$ such that a positive
minimum arises. From Lemma 8 additionally follows that both these sets are of
the form $(-\infty, \kappa_1)$ and $(\kappa_2,\infty)$, 
with $\kappa_1\leq\kappa_2$. It follows from Lemma 9 that these are both 
finite numbers.  I show that the solution corresponding to $\kappa_1$
remains positive everywhere and goes to $0$ as $\rho \to\infty$. That it is positive 
follows from the definition and the fact that $S_1$ is open, so that 
$\kappa_1\notin S_1$. Since $\kappa_1\leq\kappa_2$, we also have 
$\kappa_1\notin S_2$, so that the function must be monotonically decreasing.
Indeed, all functions corresponding to values of $\kappa$ in the interval
$[\kappa_1, \kappa_2]$ must have these properties. Thus all these functions must tend
to a limit.
We now show that no solution of (\ref{eq:a1}) can remain positive, monotonically 
decreasing, and tend to a value different from $0$. Let the asymptotic value 
of $\Psi( \rho )$ be $a$. The right-hand side of (\ref{eq:a1}) then tends to $a$, but 
$\rho\Psi^{\prime\prime}(\rho)$ must then also tend to a non-zero constant, which is 
manifestly inconsistent with the fact that $\Psi(\rho)$ is decreasing monotonically. 

All solutions corresponding to $\kappa$ values inside $[\kappa_1, \kappa_2]$
are thus positive and go to zero. Now denote the solutions
corresponding to the values $\kappa_1$ and $\kappa_2$ by $\Psi_1(\rho )$
and $\Psi_2(\rho )$ respectively.
Since they go to zero, they both behave for $\rho \to\infty$
as $\sqrt \rho K_1(2\sqrt \rho )$, see Lemma 6, and are thus integrable. 
Let us now assume that $\kappa_1<\kappa_2$.
By Lemma 8 it follows that $\Psi_1(\rho )<\Psi_2(\rho )$. Since both 
are integrable, both satisfy (\ref{eq:a3}), by Lemma 3. But two positive functions
satisfying $\Psi_1(\rho )<\Psi_2(\rho )$ cannot have the same integral. It is thus
necessary that $\kappa_1=\kappa_2$ and the solution that goes to zero
is thus unique.  The Lemma is proved, and hence the full result.

The common value of $\kappa_1=\kappa_2$
is what we have called $\kappa_0$. Its numerical value is
easily estimated through numerical integration of (\ref{eq:a1}).
We have found that $\Psi( \rho )$ becomes negative for $\kappa=1.45582$  whereas
it has a positive minimum for  $\kappa=1.45583$.
If needed, greater accuracy can be attained, but the approach is not trivial and is sketched in
Appendix \ref{app:b}. A plot of the function is provided in Figure \ref{fig:6}.

\section{Derivation of (\ref{eq:1.10}) from the equation for the scaling function}
\label{app:f}
As described in detail in \cite{ley03}, the scaling function $\Phi(x)$ generally satisfies the following
equation for all $\rho$:
\begin{equation}
\rho\int_0^\infty x^2\Phi(x)e^{-\rho x}dx=\int_0^\infty dx\,dy\,\left(
1+\frac{x}{y}
\right)
\Phi(x)\Phi(y)e^{-\rho x}\left[
1-e^{-\rho y}
\right].
\label{eq:f2}
\end{equation}
Let us now define
\begin{equation}
\Psi(\rho)=\xi\int_0^\infty\frac{\Phi(x)}{x}e^{-\rho x}dx,
\label{eq:f3}
\end{equation}
where $\xi$ is a constant we later adjust. Then (\ref{eq:f2}) becomes
\begin{equation}
-\xi\rho\Psi^{\prime\prime\prime}(\rho)=\Psi^\prime(\rho)\Psi^\prime(0)
-\left[
\Psi^\prime(\rho)
\right]^2
+\Psi^{\prime\prime}(\rho)\Psi(0)-\Psi^{\prime\prime}(\rho)
\Psi(\rho).
\label{eq:f4}
\end{equation}
This is integrated to
\begin{equation}
-\xi\rho\Psi^{\prime\prime}(\rho)=\Psi(\rho)\Psi^\prime(0)
-\Psi(\rho)
\Psi^\prime(\rho)
+\Psi^{\prime}(\rho)
\left[
\Psi(0)-\xi
\right].
\label{eq:f5}
\end{equation}
The additive constant is seen to vanish by considering the large-$\rho$ behaviour. Setting 
$\rho=0$ leads to $\xi=\Psi(0)$ and hence
\begin{equation}
\rho\Psi(0)\Psi^{\prime\prime}(\rho)=
\Psi(\rho)
\left[
\Psi^\prime(\rho)
-\Psi^\prime(0)\right]
.
\label{eq:f6}
\end{equation}
The values of $\Psi(0)$ and $\Psi^\prime(0)$ can be set equal to one by appropriate scaling, thereby
leading to (\ref{eq:1.10}). A minor point remains: (\ref{eq:f3}) does not yield the same proportionality constant 
as that found in (\ref{eq:psiscal}). This depends on the fact that the form (\ref{eq:f2}) of the integro-differential 
equation for $\Phi(x)$ implicitly assumes a definition of the typical size $S(t)$ different by a
constant factor from that used in the body of the text. 

\section{The numerical determination of $\kappa_0$, $\Gamma$, and $\Lambda$}
 \label{app:b}
 
Whereas the determination of $\kappa_0$ to the accuracy stated in Appendix \ref{app:a} is reasonably straightforward,
going any further requires some more detailed considerations. The difficulty  is that for any value 
of $\kappa\neq\kappa_0$, the distance to the exact solution grows exponentially. In order to 
obtain a solution valid up to a given distance $L$, we thus need an initial condition that is 
accurate to an accuracy of $\exp(-const./L)$. 

On the other hand, solving the equation near $\rho=0$ leads to loss of accuracy due to the vicinity of
the origin,  where the solution of 
(\ref{eq:a1}) generically diverges. The way this can be solved is to compute many terms of the Taylor series of $\Psi(\rho)$,
say 60, and to use these to compute $\Psi(\epsilon)$ for $\epsilon$ moderately small (I used $\epsilon=0.01$
and $0.005$)
to an accuracy of, say, 60 decimals. One then integrates (\ref{eq:1.10}) to very high accuracy until one 
reaches a value of $\rho$ with $\Psi^\prime(\rho)>0$ or $\Psi(\rho)<0$. We define an interval 
$[\kappa_-,\kappa_+]$, where for $\kappa_-$ $\Psi(\rho)$ becomes negative, whereas for $\kappa_+$
$\Psi^\prime(\rho)$ becomes positive. The interval is then iteratively halved util sufficient precision is reached. 
In this way we determine
\begin{equation}
\kappa_0=1.455\,824\,941\,943\,054\,763\ldots
\end{equation}
where the decimals displayed are correct.

\begin{figure}
\includegraphics{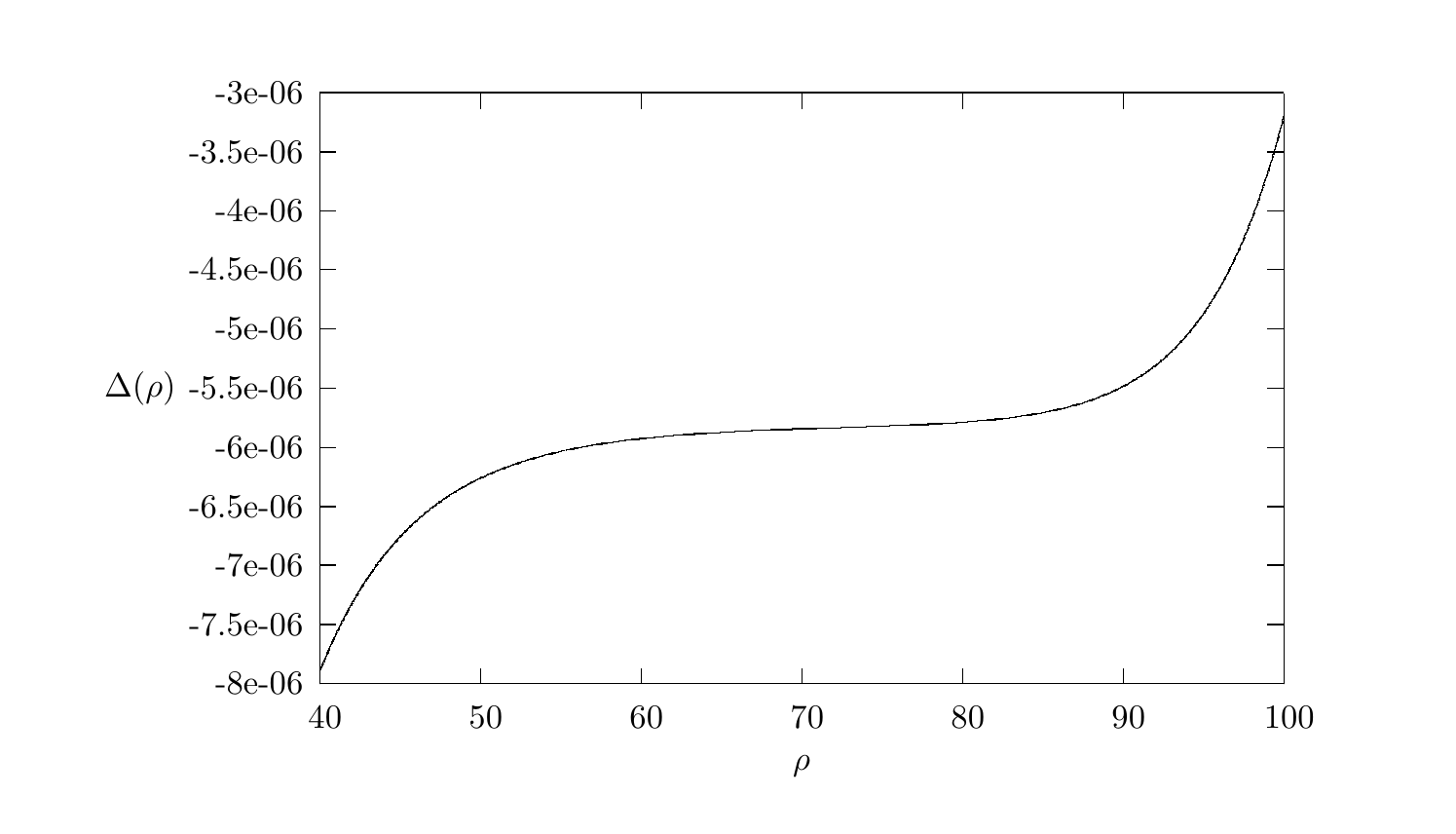}
\caption{
Ratio $\Delta(\rho)=\Psi(\rho)/[2\sqrt\rho K_1(2\sqrt\rho)]-\Gamma$, where $\Gamma=1.70787$. 
The plateau corresponds to the region of $\Psi(\rho)$ which corresponds to high accuracy
to the asymptotic region, and where the deviations at $x\gg1$ which lead eventually to 
the divergence of $\Psi(\rho)$, do not yet dominate. 
}
\label{fig:1}
\end{figure}

For the asymptotic ratio of $\Psi(\rho)$ and the asymptotic form $2\sqrt\rho K_1(2\sqrt\rho)$,
we plot this ratio minus an estimated value of $\Gamma$ given by $1.707\,87$ and show this 
in Figure \ref{fig:1}.

\begin{figure}
\includegraphics{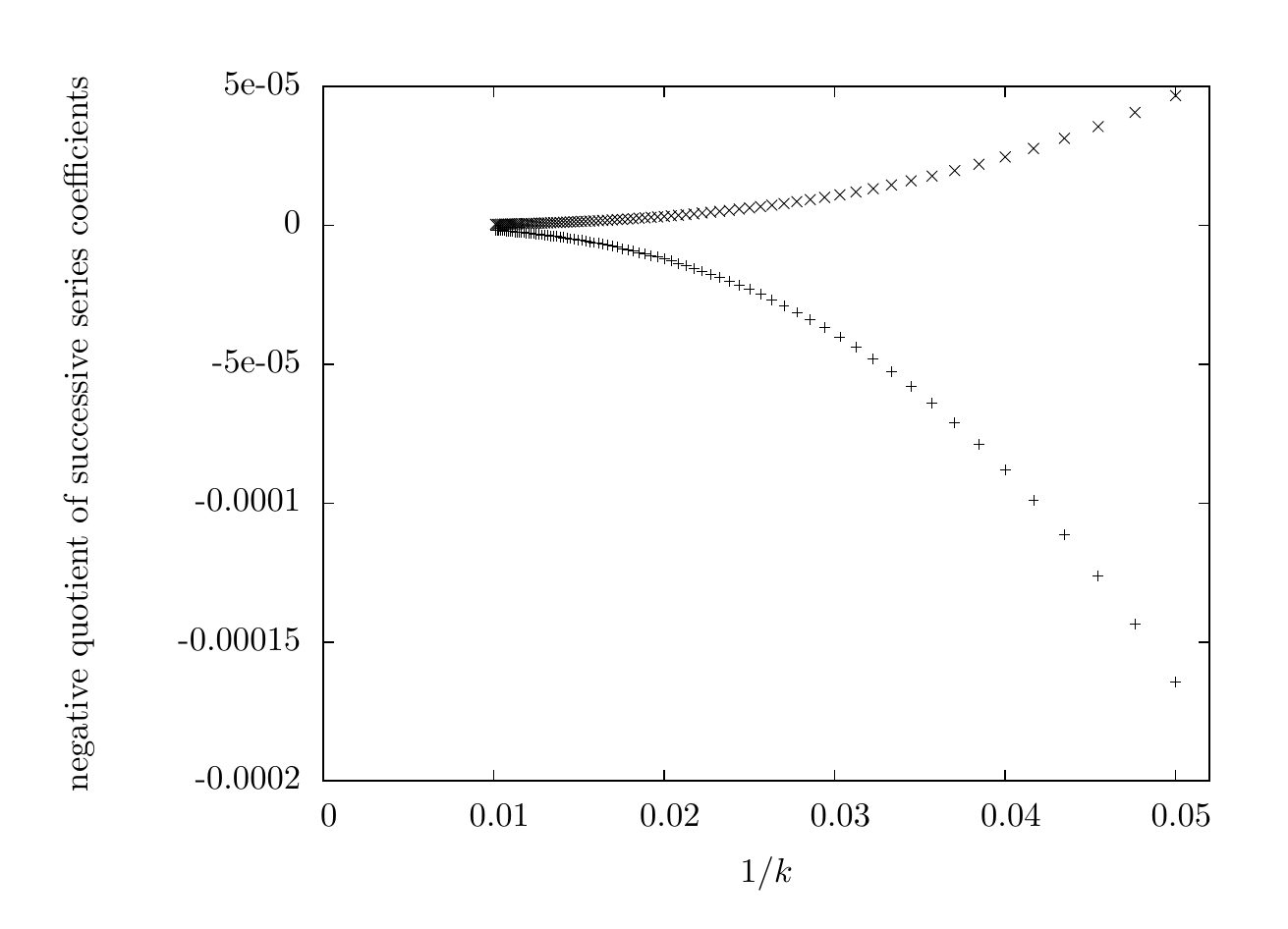}
\caption{
Ratio $-a_k/a_{k+1}$ ($+$ signs) and the corrected ratio $-b_k/b_{k+1}$ (crosses), with
$b_k=a_k(1+2/(3k^2))$, both shifted by $1.576\,132\,05$ plotted as a function of $1/k$, for  $20\leq k\leq100$. 
We note that the ratio of the $b_k$ converges more rapidly to the limiting value, indicating
that we have indeed eliminated the leading correction. Note that the shift is taken from the 
more accurate determination via Pad\'e approximants merely to show consistency. 
}
\label{fig:2}
\end{figure}

The nearest singularity of $\Psi(\rho)$ is  in leading order a simple pole, as can be seen analytically. To obtain 
an accurate numerical estimate, the simplest option is to take the first 100 coefficients of the Taylor series
\begin{equation}
\Psi(\rho)=1-\rho+\sum_{k=2}^\infty a_k \rho^{k-1}
\end{equation}
where $a_2=\kappa_0/2$, at least to good accuracy.
We then use these to estimate the radius of convergence via the ratio test. 
Since the Taylor coefficients are alternating,
the nearest singularity lies on the negative real axis. The result is shown in Figure \ref{fig:2} and appears to yield
a result of about $1.576\,13$. 

A nonlinear fit of $\ln|a_k|$ of the form $\gamma_1+\gamma_2 k+\gamma_3/k^2$ for $50\leq k\leq100$ yields
\begin{eqnarray}
\gamma_1&=&0.693\,146\,740\,47\ldots\\
\gamma_2&=&-0.454\,973\,773\,012\ldots\\
\gamma_3&=&-0.523\,738\,153\,88\ldots
\end{eqnarray}
where 11 decimals are stated, though it is clear that this is higher accuracy than warranted. 
Matching this with (\ref{eq:singdet}), one obtains $\gamma_1=\ln2$, $\gamma_2=-\ln\Lambda$,
and $\gamma_3=-\Lambda/3$. The first of these is satisfied to high accuracy. Determining $\Lambda$
using the second relation gives a result of comparable accuracy and consistent with the Pad\'e 
approximants to be discussed now, whereas the third relation is satisfied to 2 decimals, which is 
acceptable accuracy for a subleading term.

\begin{figure}
\includegraphics{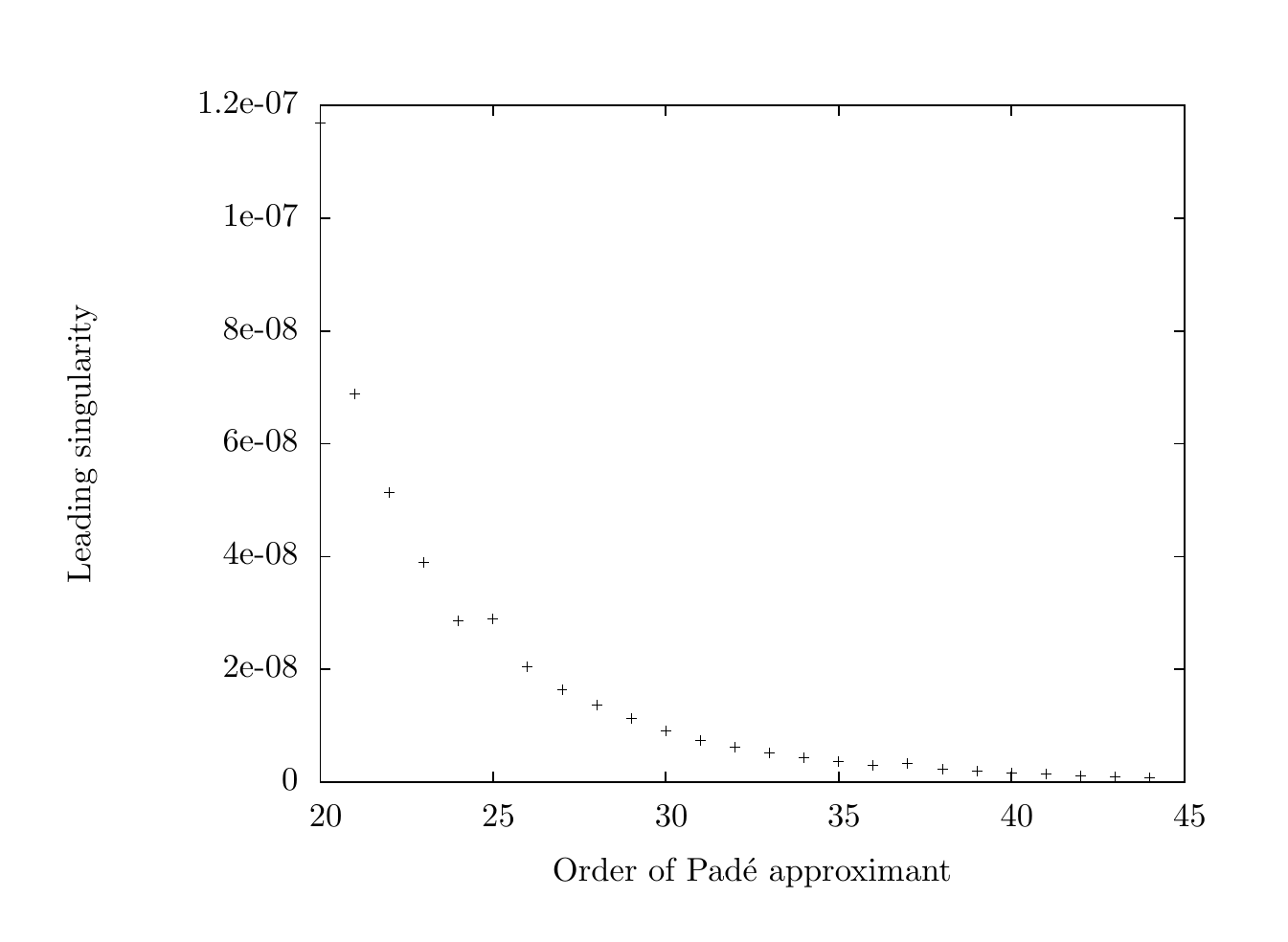}
\caption{
Zero of the denominator of the Pad\'e approximant shifted by $1.576\,132\,05$, as a 
function of the latter's order, for $20\leq k\leq45$. 
}
\label{fig:3}
\end{figure}

\begin{figure}
\includegraphics{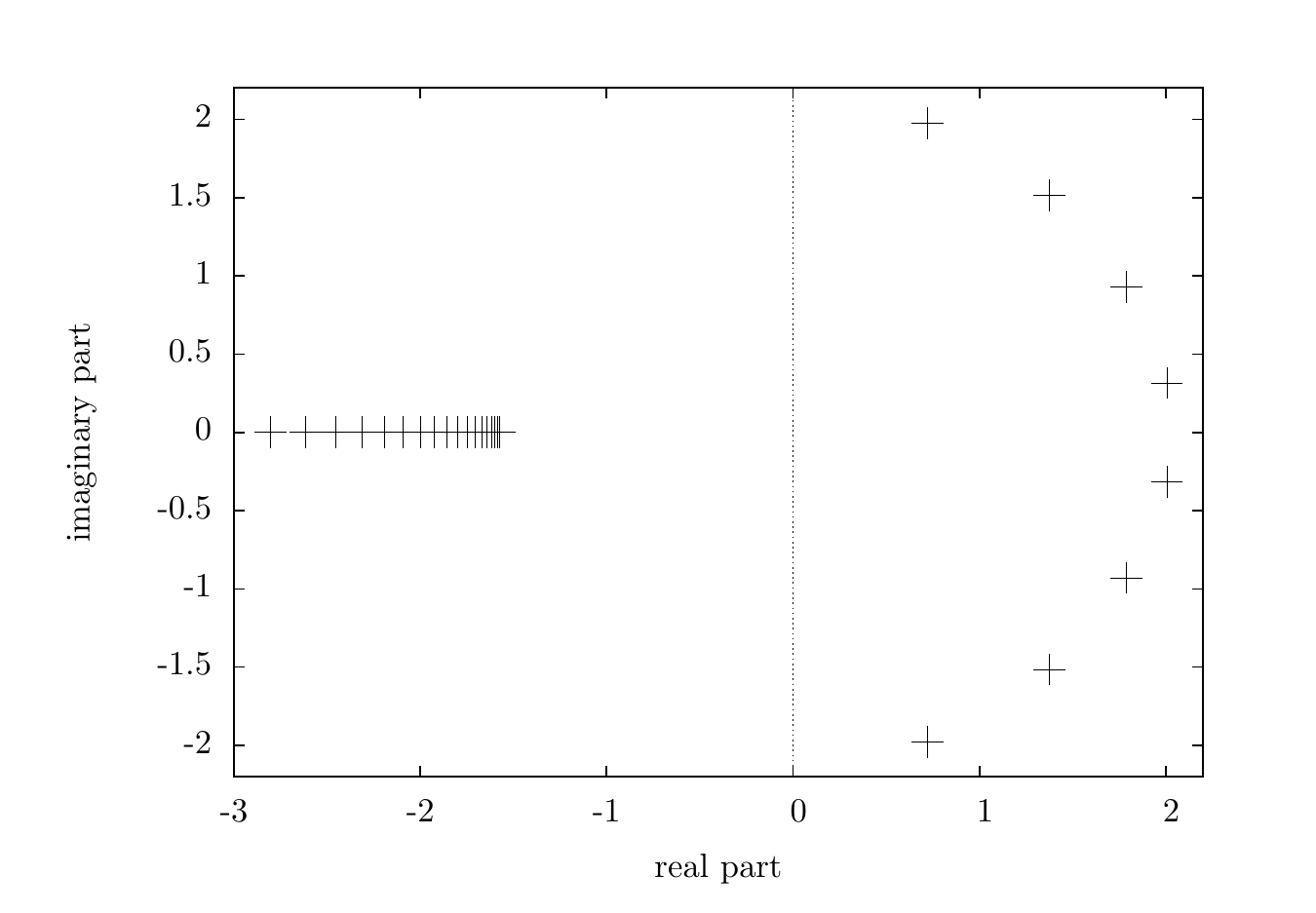}
\caption{
Zeroes of the denominator of the Pad\'e approximant of order $[45,45]$ having norm less than 3. 
The value of the nearest zero is unambiguous, and spurious zeroes do not appear, apart from one
zero (not plotted, with a value approximately $0.0146$) common to numerical accuracy to the 
numerator and the denominator, as well as several others
in the positive half-plane, which are, however all farther than $2$ from the origin. The zeroes 
accumulating near the closest zero are an indication of the existence of additional 
logarithmic singularities, as discussed in the text and shown in Appendix \ref{app:c}. 
}
\label{fig:4}
\end{figure}

As a cross-check, a Pad\'e analysis was performed. In particular, a series of diagonal
Pad\'e approximants $[M,M]$ with $20\leq M\leq45$ was generated on the Taylor
series mentioned above. It is found that there is consistently a zero of the denominator
closest to the origin, and that its value does not vary much from one approximant to the other:
we plot this in Figure \ref{fig:3}, and further show in Figure \ref{fig:4} the set of all zeroes 
of the denominator of the $[40,40]$ approximant with norm less than 3: we see that 
there are no spurious zeroes in the complex plane.
Note good agreement between this Pad\'e analysis
and the earlier ratio test.

As a final test, we use the Pad\'e approximants evaluated above to compute the residue
of the zero. The residue of a rational function $N(x)/D(x)$ at a zero $x_0$ of $D(x)$
is given by
\begin{equation}
\mathop{\mbox{\rm Res}}_{x=x_0} \frac{N(x)}{D(x)}=\frac{N(x_0)}{D^\prime(x_0)}
\end{equation}
which is readily evaluated and which, when divided by 2, yields $1.576\,1294$ for 
the approximant of order $[40,40]$. Since we had argued that 
the residue is $2\Lambda$, this is quite satisfactory agreement. 

\section{The singularity of $\Psi(\rho)$}
\label{app:c}
As has been shown in the body of this work, the function $\Psi(\rho)$ has a singularity
which is dominated by a simple pole at some point $-\Lambda$ on the negative real axis. 
Here we proceed to show that there exists a correction to the singularity, and that
 the leading and subleading behaviour of $\Psi(\rho)$ is given by
\begin{equation}
\Psi(\rho)=\left[
\frac{2\Lambda}{\rho+\Lambda}-2-\frac23
\left(
\rho+\Lambda
\right)\ln
\left(
\rho+\Lambda
\right)
\right]\left[
1+o(1)\right]
\label{eq:c1}
\end{equation}
This is shown as follows: define $y=\rho+\Lambda$. (\ref{eq:1.10}) becomes
\begin{equation}
(-\Lambda+y)\Psi^{\prime\prime}(y)-\Psi(y)\Psi^\prime(y)-\Psi(y)=0.
\label{eq:c2}
\end{equation}
We now consider the singularity near $y=0$. Matching leading singularities, we find that 
the leading behaviour is $2\Lambda/y$. To eliminate a subleading singularity of $4\Lambda/y^2$
 in the left-hand side of (\ref{eq:c2}),
we additionally need to correct this expression to $2\Lambda/y-2$.  Let us now define $v(y)$ by
\begin{equation}
v(y)=y^{-1}\left(
\Psi(y)-\frac{2\Lambda}{y}+2
\right)
\label{eq:c3}
\end{equation}
It satisfies the following equation
\begin{equation}
(\Lambda-y) \left[
y^2 v''(y)+4 y
v'(y)+2
\right]+y v(y) 
\left[y^2
v'(y)+y-2+yv(y)\right]=0.
\label{eq:c4}
\end{equation}
Clearly $v(y)$ cannot be bounded at $y=0$, since otherwise, by taking
the limit of the l.h.s.~of (\ref{eq:c4}) for $y\to0$, we obtain the contradictory
relation $2\Lambda=0$. Assuming a power law $y^p$ with $p<0$, we see that
\begin{equation}
p(p+3)=0,
\label{eq:c5}
\end{equation}
so that only $p=-3$ is  consistent, which, {\newtext however,} dominates the leading singularity and is 
thus unacceptable. This leads to assume a logarithmic divergence. Putting 
the {\em Ansatz} $v(y)=\alpha\ln y$ leads to $\alpha=-2/3$ by eliminating the leading 
singularity. A similar analysis 
to the above shows that the difference between $v(y)$ and $(-2/3)\ln y$ 
remains finite at $y=0$, from which (\ref{eq:c1}) follows. 

\section{Scaling behaviour holds for the moments}
\label{app:e}

In the following, we show that for all $n\in\mathbb{Z}$ , the quantities 
$\mu_n(t)$ and $\tilde{\mu}_{n,\epsilon}(t)$, see (\ref{eq:momdef}) and (\ref{eq:momscaldef}), 
behave identically in the limit of $\epsilon\to0$. This involves showing that the
sum
\begin{equation}
\Delta_{n,\epsilon}(t)=\sum_{j<\epsilon s}j^nc_j(s)
\label{eq:e1}
\end{equation}
is negligible as compared to $\tilde{\mu}_{n,\epsilon}(s)$, which is of order $s^{n-1}$.

The proof proceeds along somewhat different lines for $n\leq0$ and $n\geq2$. For $n=1$
the result is self-evident. We start with the latter case.

We note the fact that, since $\mu_1=1$, we always have the inequality
\begin{equation}
c_j(s)\leq\frac1j
\label{eq:e2}
\end{equation}
for all $s$. It follows that
\begin{equation}
\Delta_{n,\epsilon}(t)\leq\max_{1\leq j\leq\epsilon s}j^{n-1}\leq\left(
\epsilon s
\right)^{n-1},
\label{eq:e3}
\end{equation}
which is indeed negligible with respect to $s^{n-1}$.

For the case $n\leq0$, we first consider the case $n\leq-1$. We thus have
\begin{eqnarray}
\Delta_{n,\epsilon}(s)&=&\sum_{j=1}^Mj^nc_j(s)+\sum_{j=M+1}^\infty j^nc_j(s)
\nonumber\\
&\leq&\sum_{j=1}^Mj^n\alpha_j\exp(-s/j)+\sum_{j=M+1}^\infty j^{n-1}\\
&\leq&K_M\exp(-s/M)+CM^n
\label{eq:e4}
\end{eqnarray}
where the $\alpha_j$ are defined as in (\ref{eq:defalpha}) and $C$ is a fixed constant of order one. 
We now take for $M$ a fixed, large number such that $CM^n\leq\epsilon$.
As $s\to\infty$, the first term in the upper bound goes to zero. We thus see that, apart from a 
quantity that goes to zero as $s\to\infty$, $\Delta_{n,\epsilon}(s)$ is of order $\epsilon$ and the result is shown.

For $n=0$ the result follows from the fact that 
\begin{equation}
\dot\mu_0=-\mu_0\mu_{-1}.
\end{equation}
Since the theorem holds for all other moments, it follows for $\mu_0$. 

\section{Singularity structure of $G(z,s)$ at finite times}
\label{app:d}

Here we show that at any given fixed time, the nearest singularity of the
generating function $G(z,s)$---lying, as usual, on the negative real axis---is a pole
with a logarithmic correction of the same type as that observed as in the scaling 
function $\Psi(\rho)$. We denote this closest singularity by $-z_c(s)$

At small times the leading behaviour of the $\phi_j({s})$ is
\begin{equation}
\phi_j({s})=\lambda_j{s}^{j-1}\left[
1+O({s})
\right],
\label{eq:d1}
\end{equation}
where the $\lambda_j$ are given by (\ref{eq:beta}). This suggests introducing the following
scaling form
\begin{equation}
G(z,{s})=\frac1{s} H(z-\ln{s},{s}).
\label{eq:d2}
\end{equation}
If we introduce the new variable $x=z-\ln{s}$, the differential equation for $H(x,{s})$
becomes
\begin{equation}
H_{xx}-H_x-HH_x={s}\left(
H_{x{s}}-H
\right).
\label{eq:d3}
\end{equation}
This equation has the following amusing property: if we set the {\em Ansatz}
\begin{equation}
H(x,{s})=\sum_{m=0}^\infty {s}^mf_m(x),
\label{eq:d4}
\end{equation}
the $f_m(x)$ can be determined recursively as the solution of an ODE, which 
is nonlinear, but explicitly solvable, for $m=0$, and linear inhomogeneous
for $m\geq1$. 

The equation for $f_0(x)$ is
\begin{equation}
f_0^{\prime\prime}-f_0^\prime-f_0f_0^\prime=0. 
\label{eq:d5}
\end{equation}
The solution is given by
\begin{equation}
f_0(x)=-1-C_1 \cot
\left[
\frac{C_1(x-x_0)}{2}
\right],
\label{eq:d6}
\end{equation}
where $C_1$ and $x_0$ are integration constants. 
Without loss of generality we may put the singularity at the origin
by setting $x_0=0$ and replace the cotangent by a simple pole, setting
\begin{equation}
f_0(x)=-1-\frac{2}{x}
\label{eq:d7}
\end{equation}
Now $f_1$ solves the equation
\begin{equation}
f_1''(x)+\left(\frac{2}{x}+1\right) f_1'(x)-\frac{2
f_1(x)}{x^2}+\frac{2}{x}+1=0. 
\label{eq:d8}
\end{equation}
This has the solution
\begin{eqnarray}
f_1(x)&=&
\frac{1}{x^2} 
\left\{4
   e^{-x}\Ei(x)-
\left[P_3(x)+2
   \big[(x^2-2x+2\big) \ln x\right]+C_2\right\}
\\
P_3(x)&=&x^3-\left(C_1+3\right) x^2+2\left(
C_1+2\right) x-2 C_1
\label{eq:d9}
\end{eqnarray}
Developing around $x=0$ to third order yields
\begin{eqnarray}
f_1(x)&=&
\frac{2
C_1+C_2+4 \gamma}{x^2}-\frac{2 C_1+C_2+4
   \gamma }{x}+C_1+\frac{C_2}{2}+2
   \gamma+\nonumber\\
&&\qquad\frac{x}{18} 
   \left(-3 C_2-12 \ln x-12
   \gamma
   +4\right)+\nonumber\\
&&\qquad\frac{x^2}{72} \left(3C_2+12 \ln x+12 \gamma
-25\right)+\nonumber\\
&&\qquad\frac{x^3}{1800} \left(-15 C_2-60 \ln x-60 \gamma
+137\right)
\label{eq:d10}
\end{eqnarray}
up to terms of order 4. 
Here again $C_1$ and $C_2$ are integration constants and $\gamma$ is Euler's
constant. We cannot have an $x^{-2}$ singularity, 
so we set $C_1=-2\gamma-C_2/2$. This gives
\begin{eqnarray}
f_1(x)&=&x^3
\left(-\frac{C_2}{120}-\frac{\ln x}{30}-\frac{\gamma}{30}+\frac{137}{1800}\right)+\nonumber\\
&&\qquad x^2
\left(\frac{C_2}{24}+\frac{\ln x}{6}+\frac{\gamma}{6}-\frac{25}{72}\right)+\nonumber\\
&&\qquad x
\left(-\frac{C_2}{6}-\frac{2 \ln x}{3}-\frac{2 \gamma}{3}+\frac{2}{9}\right).
\label{eq:d11}
\end{eqnarray}
We thus find that the first order correction in ${s}$ has an $x\ln x$ correction to the leading $-2/x$
behaviour, exactly similarly to the behaviour of the scaling function $\Psi(\rho)$ near its closest 
singularity $-\Lambda$. 

We thus have as an approximate expression for $G(z,s)$ for $s$ small: close to the singularity $z_c(s)$. 
We thus have approximately:
\begin{equation}
G(z,s)=-\frac{2}{z+z_c(s)}-1-\frac{2s}{3}
\left[
z+z_c(s)
\right)]
\ln
\left[
z+z_c(s)
\right]
\label{eq:d12}
\end{equation}
and this correction, too, amounts to a correction of $-2z_c(s)/(3n^2)$ to the prefactor of the pure exponential decay, as
stated in (\ref{eq:largeclust}).
This prefactor, on the other hand, depends on constants, the value of which cannot be determined. 
Note the considerable similarity to the results obtained for the scaling function. This suggests that the scaling limit
is attained rather smoothly in the limit of large sizes.

\end{document}